\documentclass[a4paper,13.7pt]{extarticle}
\usepackage[utf8]{inputenc}
\usepackage[english]{babel}
\usepackage[T1]{fontenc}
\usepackage[a4paper,top=3cm,bottom=2cm,left=2cm,right=2cm,marginparwidth=1cm]{geometry}
\usepackage{hyperref}
\usepackage{setspace}

\usepackage{amsmath}
\usepackage{gensymb}
\usepackage[mathscr]{euscript}
\usepackage{mathtools}

\usepackage{graphicx}
\usepackage[font=footnotesize,labelfont=bf]{caption}
\usepackage{subfig}
\usepackage{wrapfig}
\usepackage[outercaption]{sidecap}

\usepackage{amssymb}
\usepackage{booktabs}
\usepackage{float}
\usepackage{color,soul}

\usepackage[parfill]{parskip}

\usepackage{natbib}
\usepackage[toc,title]{appendix}

\hypersetup{colorlinks,linkcolor={black},citecolor={blue},urlcolor={blue}}
\usepackage[english]{babel}
\graphicspath{{../}}

\usepackage{fancyhdr}
\providecommand{\keywords}[1]
{
  \small	
  \textbf{\textit{Keywords---}} #1
}
	
\usepackage{lineno}

\begin{document}
\pagenumbering{roman}

\begin{titlepage}

\begin{center}


{ \huge \bfseries Locating clustered seismicity using Distance Geometry Solvers: applications for sparse and single-borehole DAS networks.} \\[2 cm]


\textit{\\Katinka Tuinstra$^1$, Francesco Grigoli$^2$, Federica Lanza$^1$, Antonio Pio Rinaldi$^1$, \\ Andreas Fichtner$^{3}$, Stefan Wiemer$^1$}

\vspace{3cm}
\hrule
\textbf{This manuscript has been submitted for publication in Geophysical Journal International. This is a preprint submitted to arXiv, which has not undergone peer review. Subsequent versions of this manuscript might differ in content. If the manuscript is accepted, the published version will be available via a publication DOI.}
\vspace{0.15cm}
\hrule
\vspace{0.5cm}
\includegraphics[width = 20mm]{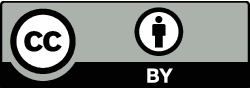}

\vspace{3cm}

\end{center}

$^1$ Swiss Seismological Service, Department of Earth Sciences, ETH Zürich, Switzerland \\
$^2$ Department of Earth Sciences, University of Pisa, Italy \\
$^3$ Institute of Geophysics, Department of Earth Sciences, ETH Zürich, Switzerland \\[1cm]

\textbf{Contact author:} katinka.tuinstra@sed.ethz.ch

\vspace{1cm}

\begin{center}
{\today}, Zürich\\[1cm]
\end{center}

\end{titlepage}
\newpage
\fancyfoot[L]{\textit{This is a non-peer reviewed preprint submitted to Geophysical Journal International}}
\vspace{10cm}
\begin{abstract}
The determination of seismic event locations with sparse networks or single-borehole systems remains a significant challenge in observational seismology. Leveraging the advantages of the location approach HADES, which was initially developed for locating clustered seismicity recorded at two stations, through the solution of a Distance Geometry Problem, we present here an improved version of the methodology: HADES-R. Where HADES previously needed a minimum of 4 absolutely located master events, HADES-R solves a least-squares problem to find the relative inter-event distances in the cluster, and uses only a single master event to find the locations of all events, and subsequently applies rotational optimiser to find the cluster orientation. It can leverage iterative station combinations if multiple receivers are available, to describe the cluster shape and orientation uncertainty with a bootstrap approach. The improved method requires P- and S-phase arrival picks, a homogeneous velocity model, a single master event with a known location, and an estimate of the cluster width. The approach is benchmarked on the 2019 Ridgecrest sequence recorded at two stations, and applied to two seismic clusters at the FORGE geothermal test site in Utah, USA, with a microseismic monitoring scenario with a DAS in a vertical borehole. Traditional procedures struggle in these settings due to the ill-posed network configuration. The azimuthal ambiguity in such a scenario is partially overcome by the assumption that all events belong to the same cluster around the master event and a cluster width estimate. We are able to find the cluster shape in both cases, although the orientation remains uncertain. HADES-R contributes to an efficient way to locate multiple events simultaneously with minimal prior information. The method's ability to constrain the cluster shape and location with only one well-located event offers promising implications, especially for environments where limited or specialised instrumentation is in use.
\end{abstract}


\newpage
\pagenumbering{arabic}

\fancyhead[HR]{Introduction}
\section{Introduction}
Microseismic monitoring plays a fundamental role in the assessment and management of the hazards and risks associated with seismicity induced by industrial activities, such as conventional and non-conventional hydrocarbon production \citep{webster2013micro, van2013microseismic, dost2017development}, underground storage of natural gas and CO$_2$ \citep{oye2013microseismic, verliac2021microseismic} and Enhanced Geothermal System (EGS) \citep{chen2019optimal, lellouch2020comparison, grigoli2022monitoring}. The first step in monitoring is detecting and locating the seismicity, which can be challenging when network coverage is poor or seismicity is occurring outside of the network design. This is for example the case in sparse seismic networks, or networks where only a single borehole is available for monitoring. In this study, we focus on locating seismicity with a single-borehole Distributed Acoustic Sensing (DAS) monitoring network using an improved relative location method.

Traditionally, a combination of permanent surface and shallow borehole sensors, as well as deep downhole geophone arrays positioned in close proximity to the seismically active area are employed in microseismic monitoring systems. For hydraulic stimulation, downhole geophones are usually lowered and clamped into dedicated boreholes, used exclusively for monitoring purposes \citep{lellouch2019observation}. More recently, the deployment of fibre-optic cables for DAS in boreholes has drawn a growing interest from the seismological community \citep{lellouchbiondi2021seismic, fichtner2023borehole} and DAS has been increasingly used as a promising and cost-effective subsurface microseismic monitoring tool \citep{daley2016field, karrenbach2019fiber,lellouch2019seismic, verdon2020microseismic, lindsey2021fiber, lellouchbiondi2021seismic}. By cementing fibre-optic cables behind the borehole casing and connecting the fiber-optic interrogator unit at the surface, seismic DAS measurements can be made during use of that same borehole, decreasing the number of required monitoring wells to be drilled. DAS offers high spatial sampling (down to 0.20 m) and high temporal resolution (up to kHz) along the fibre, providing dense and spatially coherent images of the seismic wavefield, and showing features that may have otherwise been missed by arrays with a lower spatial resolution \citep{lellouch2020comparison}. Raw DAS data can have a lower signal-to-noise ratio (SNR) when comparing individual channels against geophones. However, the deep deployment capabilities of downhole DAS, with the SNR increasing steadily as the antropogenic noise from surface activities decreases, enhances the capability to detect and locate microseismicity. An important limitation of DAS is that it measures only along the axial direction of the fiber. In a borehole setting, this results in only single-component measurements that give no azimuthal information for recorded events. Hence, locating seismicity in such a case requires additional instrumentation or assumptions, which we try to address in this work.

An earthquake location is most commonly determined by comparing seismic-phase arrival times  observed at seismic stations at the surface with associated uncertainties with arrival times predicted for a given seismic velocity model. Absolute location methods based on arrival times were introduced by \citet{geiger1912probability}, and nowadays constitute one of the most used methods for earthquake location (e.g., NonLinLoc (NLL), by \citet{lomax2009earthquake, lomax2020absolute}. These absolute location methods are often accurate but have limitations. They depend heavily on the (often unknown) velocity model and on the azimuthal coverage of the seismic network, which, in underground settings, is commonly limited due to logistical constraints. As a consequence, in single-borehole monitoring configurations, location methods remain challenging for both DAS \citep[e.g., ][]{karrenbach2017hydraulic, lellouch2021low}, and geophones \citep[e.g., ][]{maxwell2009microseismic,eisner2010comparison,eisner2011challenges}. Sparse arrays of geophones or seismic stations often have trouble to uniquely locate seismicity, especially when it is occurring 'out-of-network'. Absolute location then depends on analysis of three components \citep[e.g., ][]{roberts1989real}, which is not always possible due to noise, weak events, or single-component sensors. Although methods designed for downhole geophones in a single borehole exist  \citep{eaton_2018, abercrombie1995earthquake}, they require polarisation (hodogram) analysis to determine the arrival direction of a wave to provide an estimate of the source backazimuth, a procedure that is often based on manual intervention and ad-hoc parametrisation. Given the use of both vertical and horizontal components, these methods are not applicable to DAS, which measures the strain only in the axial direction of the fibre. While there is total azimuthal ambiguity for a single borehole instrumented with fibre-optic cable, a location based on distance and depth is possible, e.g., in the beamforming approach used by \citet{lellouch2020comparison}.

This presents a need to overcome challenges due to poor network coverage, single-component measurements, or unknown velocity models. In relative (re-)location methods, seismic events are located with respect to one another. An advantage of this approach is that the locations are less dependent on the seismic velocity model, as only the velocity between the events is taken into account. Double-difference algorithms, based on cross-correlations of the seismic events at each station (i.e., similarity of the ray-paths between events-station pairs) are often embedded in relative location methods. Examples of relative locations algorithms are HypoDD \citep{waldhauser2000double}, GrowClust by \citet{trugman2017growclust}, NLL-SSST (NonLinLoc with Smooth, Source-specific, Station Travel-time) by \citet{lomax2022high}, and HADES by \citet{grigoli2021relative}, to name a few. Thus, relative relocation methods can prove advantageous in microseismic monitoring of clustered seismicity when many events are expected to occur in a short amount of time in a small clustered volume \citep{stark1992microearthquakes}. 

Building on the advantages of relative location principles and utilising the numerous channels available in DAS data, we propose a methodology to locate clustered microseismic events. The approach targets both sparse conventional surface networks and single-borehole configurations, with the goal of reducing the azimuthal ambiguity in clustered locations. This addresses a challenge often faced in these common but ill-posed network settings. An important prior assumption is that the seismicity to be located is part of the same cluster, which could be tested with waveform clustering \citep{menke1999using, baisch2008earthquake} prior to locating. The method builds on the work of \citet{grigoli2021relative}, which proposed an efficient method based on Distance Geometry Problem (DGP) solvers: HADES (eartHquake locAtion via Distance gEometry Solvers). HADES is based on protein structure determination problems \citep{sit2009geometric} which have been adapted to locate clustered seismicity with sparse surface networks of one or two stations. In this paper, we further extend the method to be used with DAS or array data, and reduce the minimum required a-priori known absolute locations from four to a single event (hereafter, the \textit{master event}). This is made possible by creating a relative \textit{reference frame} of the master event, and four other events with unknown location, and subsequently locating all other seismicity with respect to this frame.  The master event could be, e.g., a check shot, or the largest event in the sequence that is also recorded by the surface array, and that can therefore be located with high confidence. We perform rotational optimisation of the relatively located cluster around the master event to find the best-fitting orientation, and apply a bootstrap approach by iterating over numerous channel combinations in dense arrays or DAS to form an estimate of shape- and orientation uncertainty. The required inputs for our proposed new locator HADES-R (HADES-Relative) consist of only (1) a minimum of one master event, and (2)  P- and S- arrival times. As a result, any type of measurement can be used, eliminating the complexity arising from the different units of measurements between DAS (strain-rate) and conventional geophones (velocity or displacement), and allowing the implementation of mixed sensor networks.

HADES-R benefits a specific niche of seismic location problems: clustered seismicity that is occurring out-of-network and is either recorded by very ill-posed or sparse networks. This allows us to monitor the evolution of seismicity within a volume of interest, when only a minimal seismic network is operational, for example a single borehole DAS. Since the HADES method was originally designed for networks of one or two stations, it was inherently designed for locating seismicity occurring out-of-network. This work does therefore not only benefit the downhole DAS community for locating (micro)seismicity, but also enables us to locate seismic clusters recorded by conventional sparse surface networks with very little a-priori information needed. 

In the following sections, we explain the methodology of the new HADES-R approach, and benchmark it on the Ridgecrest seismic sequence in California. We then show synthetic test results for the downhole DAS configuration that is the same as the later real-data examples. The method is then applied to two real microseismic datasets recorded at the Utah Frontier Observatory for Research in Geothermal Energy (FORGE) site during the pilot hydraulic stimulation of 2019. Finally, we compare our results against other location results obtained with DAS by \citet{lellouch2021low}.

\rhead{Methods}
\section{Methodology: Seismic cluster determination using Distance Geometry Solvers}

The HADES location algorithm is based on the solution of a Distance Geometry Problem: the determination of the coordinates of a set of points in space by knowing only the mutual distance between each pair of points. Given known exact distances between all points, their position in a cluster can be uniquely found \citep{liberti2014euclidean}. When only the approximated inter-event distance is known, the absolute location of the points will not be exact, but it can still be estimated. The method relies on the Molecular Geometric Build-Up technique \citep{sit2009geometric, wu2007updated}, used in biochemistry to determine the structures of proteins. HADES, as described in \cite{grigoli2021relative}, applies this technique to solve an earthquake location problem in sparse networks of one or two stations, where absolute techniques fail due to either lack of horizontal components, or too much noise to estimate the back-azimuth. The primary assumption is that all the seismic events to be located belong to the same cluster, and that the overall size of the cluster is much smaller than the distance between the cluster and the receivers. Input data are: (1) a previously well-located event (hereafter, the \textit{master event}), (2) a homogeneous velocity model, (3) an estimate of the width of the cluster, and (4) the arrival time picks of the P- and S-wave arrivals which are used to calculate the approximate inter-event distances between all event pairs. In a seismic cluster, the Euclidean distance $\left\|\mathbf{r}_{a b}^{S1}\right\|$ between two events $a$ and $b$ can be estimated from the double difference in arrival times of S- and P- phases ($t_s -$ $t_p$) measured at a single seismic station $S1$,

\begin{equation}
\left\|\mathbf{r}_{a b}^{S1}\right\|=k_{\mathrm{v}}\left|\left(t_{\mathrm{s}}^{a}-t_{\mathrm{p}}^{a}\right)-\left(t_{\mathrm{s}}^{b}-t_{\mathrm{p}}^{b}\right)\right|,
\end{equation}

with $k_{\mathrm{v}} = v_{\mathrm{P}} v_{\mathrm{S}} /\left(v_{\mathrm{P}}-v_{\mathrm{S}}\right)$ depending on homogeneous P- and S-wave speed estimates $v_p$ and $v_s$. The distance  will be mathematically exact if the two events and the station are located along the same line, otherwise it will be approximated (see Appendix \ref{sec:sup_iedist}, or \cite{grigoli2021relative} for a detailed description). For two stations ($S1$ and $S2$) the inter-event distance $\mathbf{r}^{total}_{a b}$ between two events can be better approximated by combining the inter-event distance estimated at each station as follows:

\begin{equation}
\left\|\mathbf{r}^{total}_{a b}\right\| \approx \sqrt{\left\|\mathbf{r}_{a b}^{S1}\right\|^{2}+\left\|\mathbf{r}_{a b}^{S2}\right\|^{2}}. 
\end{equation}

\begin{figure}[H]
    \centering
    \includegraphics[width=\textwidth]{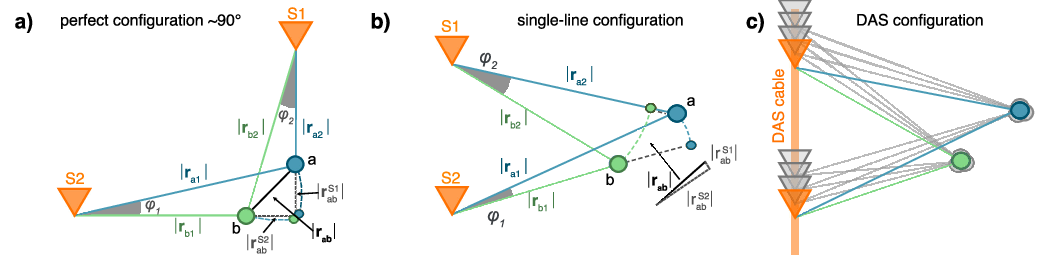}
    \caption{Three angular configurations of receiver pairs (S1, S2, inverted orange triangles) and two events true locations (a, b, large coloured circles) and their projections as seen by single stations (small coloured circles). a) Perfectly orthogonal configuration with respect to the seismicity cluster leads to better estimation of the inter-event distance using the Pythagorean theorem. $\left\|\mathbf{r}^{S1}_{ab}\right\|$ and $\left\|\mathbf{r}^{S2}_{ab}\right\|$ are the estimated inter-event distances using only one station, and, when combined, they can be used to calculate the approximate the inter-event distance $\left\|\mathbf{r}_{ab}\right\|$. b) The two stations are not orthogonal with respect to the two events. This leads to poor estimations of the inter-event distance $\left\|\mathbf{r}\right\|$. The black triangle shows the estimation of the inter-event distance between event a and b. c) The configuration with an iteration over channel pairs with DAS or other dense arrays, where the different configurations could result in slightly different event locations. Modified after \citet{grigoli2021relative}.}
    \label{fig:ang_conf}
\end{figure}

With two stations, the approximation is closer to the exact solution if the stations are in an orthogonal position with respect to the geometric centre of the seismic cluster (see Fig. \ref{fig:ang_conf}a), and if all events and stations lie within the same plane. Since in real cases it is almost impossible to fulfil this requirement, what we generally obtain is an approximated inter-event distance between two events, forming a projection on the plane passing through the stations and one of the two events. According to \cite{grigoli2021relative}, the following conditions should be met for an optimal inter-event distance approximation: (1) the distance between the station(s) and the centroid of the cluster is much larger than the inter-event distance (i.e., the angle between the station and the outer events of the cluster $\phi \rightarrow 0$, see Fig. \ref{fig:ang_conf}), (2) the axes between the receivers and the geometric centre of the cluster are close to orthogonal, and (3) the master event is non-coplanar with respect to the other events that form the reference frame (this will be explained later in this section). The first and most important condition means that the seismicity forms a single cluster, which is much smaller compared to the average source-station distance. This condition can be tested in advance using waveform clustering techniques \citep[e.g., ][]{menke1999using, baisch2008earthquake}, where waveform cross-correlation is taken as a measure of distance between events. Fig. \ref{fig:ang_conf}b shows how a non-orthogonal configuration of two stations and the seismic cluster can lead to increasingly poor estimates of the inter-event distance, as is the case with DAS in a vertical borehole (Fig. \ref{fig:ang_conf}c). It is important to mention that we only described a simple way to obtain approximated inter-event distances. In theory, HADES can perfectly reconstruct the shape of the cluster if exact inter-event distances are provided as input. 

In the method developed by \cite{grigoli2021relative} (HADES), at least four master events with known locations are required to find the absolute location of the cluster. These master events are the absolute 'anchors' of the relative seismic cluster to its absolute location. Any rotational error of the cluster is minimised in the horizontal (latitude-longitude) plane, whereas rotational errors in depth may remain. The dependency of the original HADES on at least four master events with accurate locations leads to two limitations: firstly, some sequences might have less than four well-located master events, and secondly, if the location uncertainties of the master events are large, these uncertainties will propagate into the reference coordinate system and force the cluster into a potentially incorrect shape. In HADES-R, we extend HADES by reducing the minimum number of required master events from four to just one event. If many receiver points are available, such as with DAS, we also extend the method by running many iterations over different channel pairs, in order to obtain an estimate of the location uncertainties.

\begin{figure}[H]
    \centering
    \includegraphics[width=0.6\textwidth]{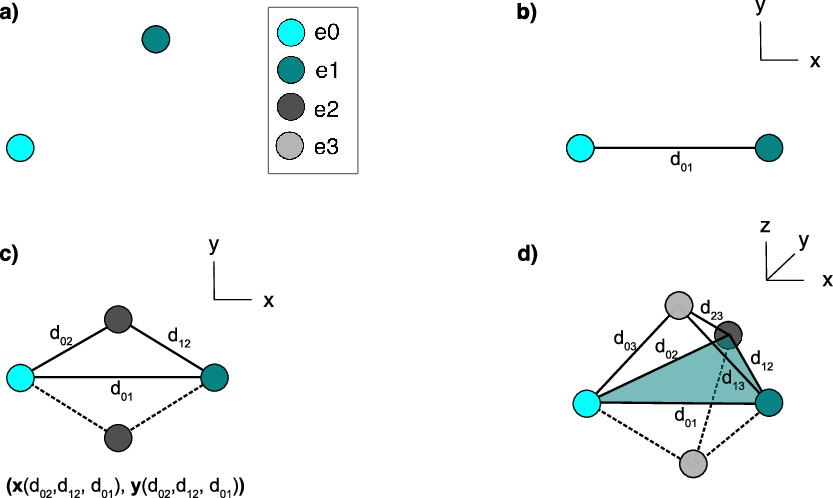}
    \caption{Construction of a relative coordinate frame using only one master event (Event 0). A) the master event (Event 0) is placed at the origin of our relative reference frame $(x_{0},y_{0},z_{0})=(0,0,0)$. Then, B) The distance ($d_{01}$) between Event 0 and the first reference event Event 1 is calculated to set the coordinates of Event 1: its location is $(x_{0},y_{0},z_{0})=(d_{01},0,0)$. C) The second reference event (Event 2) is located following the same approach as in B). Event 2 can be arbitrarily located on the positive or negative side of the line between event 0 and event 1 to form a plane. D) the third reference event can be positioned under or above the 2D plane in c). The distance of event 3 to the plane is an estimate that will dictate the cluster width.}
    \label{fig:methods_refframe}
\end{figure}

Using the master event, a relative reference frame can be built by adding three other \textit{reference events}. It is not necessary to know the absolute location of the reference events, however an estimate of the width of the reference frame is required. If additional information is available, we recommend choosing reference events that are non-coplanar and are spread throughout the cluster. This ensures a better approximation of the cluster shape. To set up the relative reference frame, we follow the geometrical build-up algorithm (we show the procedure in Fig. \ref{fig:methods_refframe}). To highlight that these are the reference events, we use the symbol $d$ for the distances between the reference frame events, instead of the $\left\|r\right\|$ used for inter-event distances. The method consists of placing the master event $e_0$ at the origin of an arbitrary coordinate system ($(x_{0}, y_{0}, z_{0})$ = (0,0,0)), and then the estimated distance between $e_0$ and the first reference event $e_1$ ($d_{01}$) is used to place $e_1$ along the x-axis. These positions are only \textit{relative positions} in an arbitrary space. They do not represent the final location coordinates. The position of the second reference event in $(x_{2}, y_{2}, z_{2})$ forms the reference plane ($e_0, e_1, e_2$), and is given by: 

\begin{equation}
\begin{aligned}
    &x_{2} = (d^2_{02}-d^2_{12})/2x_{1} + x_{1}/2 ,\\
    &y_{2} = \pm (d^2_{02} - x_{2}^2)^{1/2} ,\\
    &z_{2} = 0 ,
\end{aligned}
\end{equation}

in this way, we construct a 2D reference frame that lies on the plane $z=0$ (Fig. \ref{fig:ang_conf}c), and the rest of the events in the cluster can be located relatively to this frame. Finally, a third reference event $e_3$ is placed above or below the reference plane that we saw in the 2D version, see Fig. \ref{fig:methods_refframe}. Its position $(x_{3}, y_{3}, z_{3})$ can be exactly obtained with:
 
 \begin{equation}
 \begin{aligned}
     &x_3 = (d_{03}^2 - d_{13}^2)/(2x_{1}) + (x_{1})/2 , \\
     &y_3 = (d^2_{13}-d_{32}^2-(x_{3}-x_{1})^2+(x_{3}-x_{2})^2)/2y_{2} + y_{2}/2 ,  \\
     &z_3 = \pm(d_{12}^2-x_{3}^2-y_{3}^2)^{1/2} .
 \end{aligned}
 \end{equation}

It is important to note that the use of approximated inter-event distances does not guarantee that a real-valued, non-trivial solution to the last equation exists (i.e., the square root can be complex). In general, $z_3$ will collapse to 0 if calculated using the approximated inter-event distances based on the differential arrival times, as is the case in this paper. Hence, we need additional information to estimate the $z_3$ reference events, or have an initial idea of the three-dimensional distance of the third event. This estimate may come from the aforementioned clustering analysis, or other prior knowledge of expected cluster bounds (i.e., geological constraints). The $z_3$ can thus be manually set as a single value or a range of widths can be used. This is why HADES-R does not truly overcome the azimuthal problem, but only tries to approximate it given a limited 'width' of the reference frame and cluster. Both positions of $e_2$ (Fig. \ref{fig:methods_refframe}c) are equivalent, and so is the position of $e_3$ \ref{fig:methods_refframe}d) above or below the plane $(e_0, e_1, e_2)$. We account for this ambiguity later in the rotational post-processing component, by also iterating over multiple mirrored variations of the final cluster.

The obtained reference frame will be used to locate other events. This process is the same as in the original version of HADES, and for a more mathematical and graphical explanation we refer to Appendix \ref{sec:sup_clusloc} and \cite{grigoli2021relative}. Given a seismic cluster with $k$ events at hypocentral coordinates $\mathbf{x}_{1},\mathbf{x}_{2},...,\mathbf{x}_{k}$ of each event, we would like to locate a new event $k + 1$. Each new event $k + 1$ is placed at the origin of a reference frame $\mathbf{x}_{k+1}=(0,0,0)$, and the already-located events $1, ..., k$ are located relative to it via rotations and translations, such that the misfit between the initial and the new locations is minimised in a least-squares approach. This process is repeated for each new event that needs to be located, meaning that all events are relocated multiple times in the procedure. This approach is applied iteratively for all events in the cluster..

Due to the lack of azimuthal or 'cluster width' information, the single master-event method HADES-R outputs a slightly planar version of the cluster with respect to the original version of HADES, especially in non-ideal station configurations (Fig. \ref{fig:ang_conf}). The placement of the second and third reference event on the positive or negative side of the reference system determines the handedness of the system. Both left- and right-handedness can be tested to see which system produces the minimal misfit. Hence, HADES-R can reconstruct the shape of the cluster, but it has limitations in resolving the absolute location of the events within the cluster.

\begin{figure}[H]
    \centering
    \includegraphics[width=0.6\textwidth]{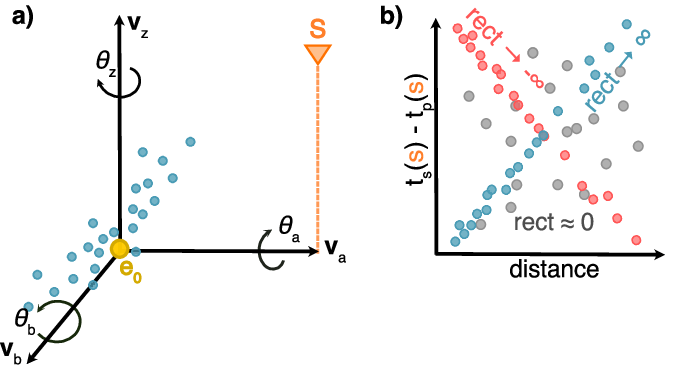}
    \caption{a) Cluster rotation around axis vectors $\mathbf{v}_z, \mathbf{v}_a$, $\mathbf{v}_b$. The master event $e_0$ is at the origin of the system shown as a solid yellow circle, and the other events of a hypothetical cluster are indicated with blue dots. b) Rectilinearity example. If the distribution between the distance of the events to the station versus the differential arrival times $(t_s-t_p)$ is positive linear (blue dots), this is physical and correct. If it forms a negative linear relationship this is non-physical (red dots) and the cluster is likely rotated in the opposite direction. If the distribution is random (grey dots) then either the rotation is incorrect, or the shape of the cluster has been poorly reconstructed.}
    \label{fig:methods_quat}
\end{figure}

While the events have now been located relative to each other, the determination of their absolute spatial location is pending. In the original HADES version, the minimum four master events were used to constrain the cluster to its absolute location. In the proposed HADES-R, only the absolute location of the single master event is used. This means that the correct rotation of the cluster to its absolute location cannot be found directly. To this end, we developed a cluster rotation tool, that employs quaternions \citep{shoemake1985animating}, to perform three-dimensional rotations around the master event (See Appendix \ref{sec:sup_quat}). It searches for the correct distribution of observed arrival times versus distance to the stations. In detail, to orient a seismic cluster, we consider the master event as the rotational origin of the seismic cluster. The three rotation optimisation axes are then $\mathbf{v}_z$: a vertical axis from the origin pointing upwards to the surface, $\mathbf{v}_a$: an orthogonal axis from the origin pointing to one selected station on the horizontally projected plane, and $\mathbf{v}_b$: the cross product of the first two axes. The choice of station that governs $\mathbf{v}_a$ is interchangeable. The three unit vectors ($\mathbf{v_z}, \mathbf{v_a}, \mathbf{v_b}$) are computed for these three axes (See Fig. \ref{fig:methods_quat}a).

To show how close the rotation of the cluster is to the true position of the events, without having to calculate synthetic arrival times with a potentially incorrect velocity model, we use a metric called rectilinearity (Fig. \ref{fig:methods_quat}b). Rectilinearity measures how close a distribution is to a linear relationship and it is a form of dimensionality reduction of the arrival time and distance data, that is computed via Principal Component Analysis (PCA) \citep{jolliffe2002principal}. We obtain the rectilinearity by taking a data matrix $\mathbf{X}$ with the sorted and demeaned distances $\mathbf{r}$ and arrival time differences $\mathbf{D}$ of all $N$ events, 

\begin{equation}
    \mathbf{X} = 
    \begin{bmatrix}
    r_1 & D_1 \\
    \vdots & \vdots \\
    r_N & D_N,
    \end{bmatrix}
\end{equation}

From the transpose of $\mathbf{X}$ we can compute the covariance matrix:

\begin{equation}
    \mathbf{C} = \text{cov}(\mathbf{X}^T).
\end{equation}

The rectilinearity is then defined as the ratio between the maximum and minimum normalised eigenvalues ($\lambda_{max}$ and $\lambda_{min}$) of the data covariance matrix: 

\begin{equation}
    \mathcal{R} = \frac{\lambda_{max}}{\lambda_{min}}.
\end{equation}

A perfect linear relationship between differential arrival times and distance to the receiver produces vanishing eigenvalues except for the first, and therefore infinite rectilinearity.

The optimal rotation angle for each of the three axes ($\mathbf{\theta}_a, \mathbf{\theta}_b, \mathbf{\theta}_c$) is the one that maximises the rectilinearity of the absolute distance between the receiver and the events, with respect to the differential arrival time $t_s-t_p$ at each receiver (Fig. \ref{fig:methods_quat}b, blue dots). Ideally, the distribution of $t_s-t_p$ is linearly increasing with increasing distance to the receivers. A cluster in which the differential arrival times to distance ratio decreases is non-physical and thus inversely rotated (Fig. \ref{fig:methods_quat}b, red dots). A non-linear relationship of the $t_s-t_p$ versus distance can be caused by rotation errors, poor cluster shape reconstruction or large uncertainties in the picks (Fig. \ref{fig:methods_quat}b, grey dots). The inverse of the rectilinearity $1/\mathcal{R}$ can be used as a misfit function that has to be minimised. A basin hopping global optimiser is used to search for the optimal angle around each axis ($\mathbf{\theta}_a, \mathbf{\theta}_b, \mathbf{\theta}_c$). The basin hopping algorithm was developed for finding the positions of atoms in molecules \citep{wales1997global}: it randomly chooses coordinates in the misfit function, performs a local minimisation, and based on the minimal value chooses via an acceptance criterion whether or not to 'hop' to new random coordinates to repeat the process. This should help finding the global minimum in a high-relief misfit function with many minima. With some angular station configurations, the solution to the rotation problem may be non-unique, i.e., multiple rotations of the cluster will result in equivalent distance-arrival time distributions. This can occur in a situation with a single vertical borehole (Fig. \ref{fig:ang_conf}), where lateral rotations of events could result in equivalent arrival times. In such ill-configured cases, HADES-R may not be able to fully resolve the azimuthal ambiguity. This complexity with angular station configurations is thus the core reason for the development of the aforementioned cluster rotation tool, as it can easily and efficiently assist in finding the global minimum in a high-relief misfit function with many minima. In applying HADES-R to DAS data, the numerous channels available for picking arrivals allow for the use of an iterative bootstrap approach. See Fig. \ref{fig:ang_conf}c for an illustration. This method aids in constraining uncertainties and identifying possible biases or ambiguities in azimuthal orientation. In the result section, we show two-dimensional histograms of the output of 100 iterations of HADES-R with rotation optimisation conducted over 100 channel pairs. The criterion for choosing channel pairs is that they have to lie as far apart as possible to ensure the best achievable network aperture and thus angular configuration (Fig. \ref{fig:ang_conf}). As the data section will explain, only a portion of the DAS cable was suitable for picking the P- and S-phase arrivals. The number of iterations (100) is based on the maximum amount of channel pairs, while still ensuring maximal aperture conditions and comparability between results from different tests. The best-estimate location in each plot is the location estimate with the highest rectilinearity, that also allows to examine the inferred cluster shape. 

\rhead{Benchmark Test}
\section{Benchmark test on the 2019 Ridgecrest (California) seismic sequence}

\begin{figure}[H]
    \centering
    \includegraphics[width=0.9\textwidth]{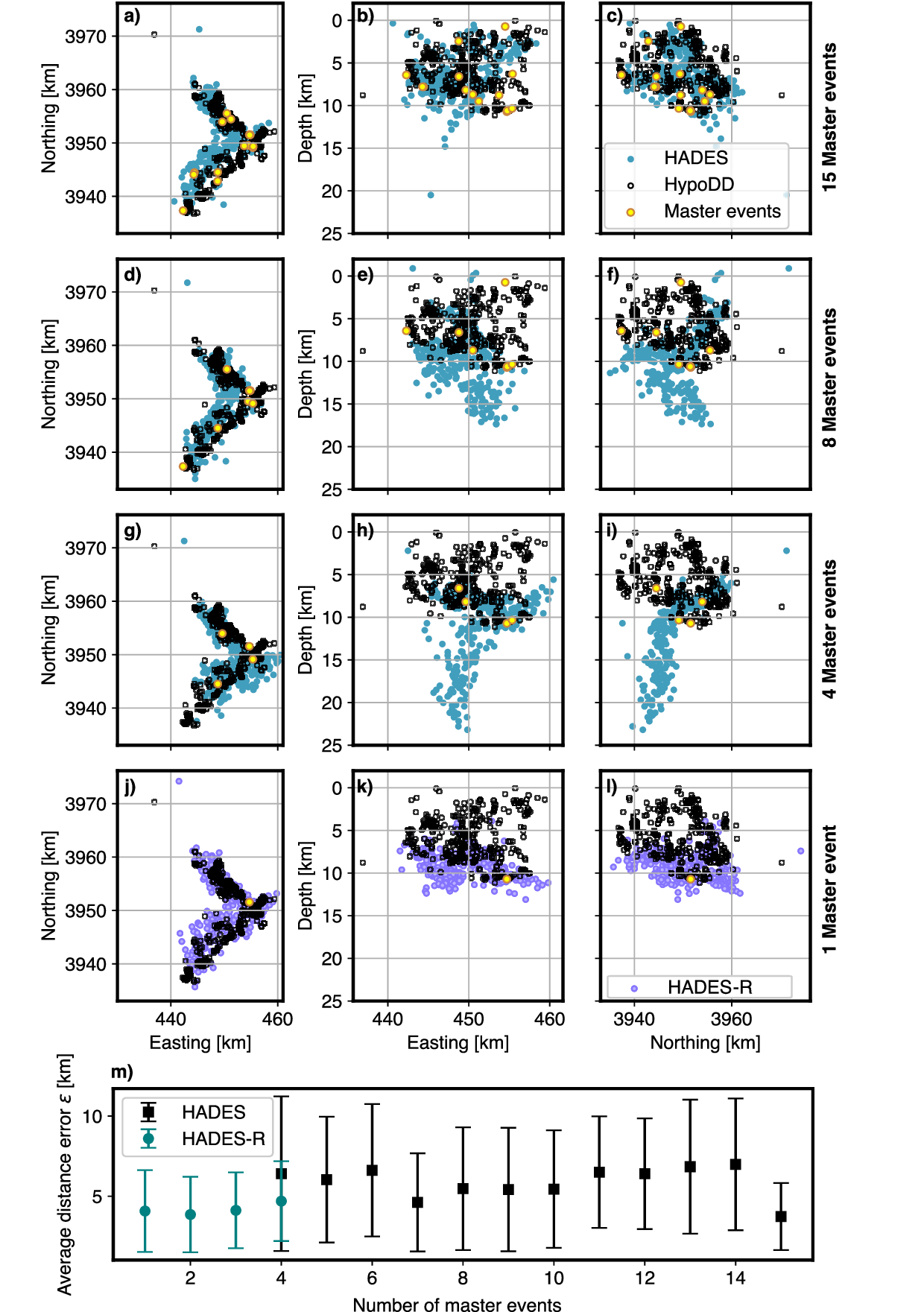}
    \caption{Comparison of the Ridgecrest results with a decreasing number of master events HADES (a-i) and finally the HADES-R (j, k, l) with one master event, against the HypoDD locations by \cite{liu2020rapid}. The first three rows are modified after results from \cite{grigoli2021relative}. (m) comparison of the average distance error and their standard deviations (error bars) between HypoDD and HADES or HADES-R, for an increasing number of master events. For the HADES-R results, scenarios for two, three, and four master events represent the average of two, three and four runs, respectively, in which each run involves a different single master event. HADES results are all single runs with increasing master events.}
    \label{fig:ridge_all}
\end{figure}

HADES-R is benchmarked with the improvements developed in this study using the 2019 Ridgecrest seismic sequence, in California, USA. This is the same dataset used by \cite{grigoli2021relative} as a real data application for validating the original HADES. The 2019 Ridgecrest seismic sequence initiated on 4 July 2019 with an $M_w$ 6.4 earthquake and it was followed two days later by an $M_w$ 7.1 earthquake on 6 July 2019 \citep{liu2020rapid}. The $M_w$ 6.4 event generated a prolific aftershock sequence on 4 and 5 July 2019, activating an L-shaped fault system \citep{ross2019hierarchical}. Similarly to \cite{grigoli2021relative}, we used the same subset of 320 events from a double-difference catalogue generated by \citet{liu2020rapid}, and we considered the same station configuration of two near-orthogonal stations at approximately 60 km distance, while decreasing the amount of master events. In Fig. \ref{fig:ridge_all} we compare the performance of HADES (teal dots)  for different cases, in which the number of master events decreases from 15, 8, and 4 to one master event, and HADES-R (purple dots). The locations obtained with both HADES versions are compared against the HypoDD locations obtained by \citet{liu2020rapid} for which a minimum of 30 stations were used per event (open black circles in Fig. 4). For HADES, we observe that the reconstruction of the shape and orientation of the cluster decreases in quality when fewer than 15 master events are used (upper three rows in Fig. 4). For HADES-R, with one master event, we note how the cluster shape is clearly aligning in the horizontal (Easting-Northing) plane, and although the depth distribution might be slightly more planar than the HypoDD events, they are more in the same range than the HADES results with 4 and 8 master events. In the lower panel (Fig. \ref{fig:ridge_all}m), we conduct a performance comparison between HADES-R and HADES. We calculate the average location distance error and standard deviation between HypoDD solutions and both HADES and HADES-R, considering an increasing number of master events. For HADES, a master event is added from the minimum of 4 up to 15. In HADES-R, only one master event can be employed in each run. To simulate the performance of using multiple master events, we run HADES-R multiple times, each time changing the master event. The resulting locations are the average computed across all runs. This procedure is repeated for scenarios involving four, three and two master events. It's worth noting that for the case with a single master event, no averaging is performed as this is simply the result of one run of HADES-R. The master events of the HADES-R runs correspond to the four master events in HADES. This test shows that HADES-R has a comparable  average error and spread to the HADES run with 15 master events. This implies that when very few master events are available, or their location is uncertain, it may be preferable to run HADES-R multiple times for different single master events. This approach can be used instead of, or in addition to, running HADES. 

\begin{figure}[H]
    \centering
    \includegraphics[width=0.8\textwidth]{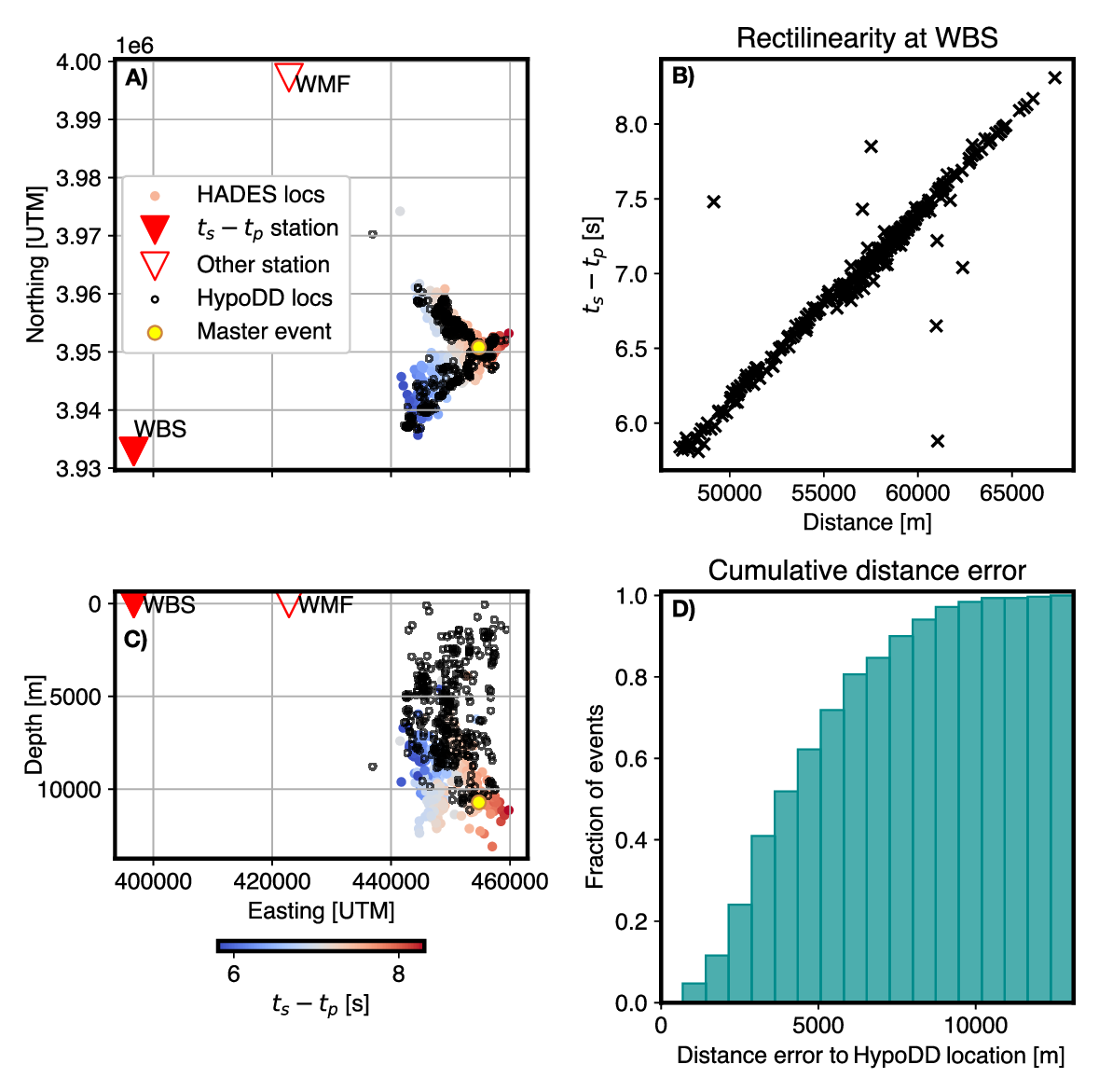}
    \caption{Final rotated locations of Ridgecrest aftershock sequence obtained with HADES-R. a) Plane view, and c) cross-section of the results after automated rotation optimisation. Events are colour-coded according to the differential arrival time (reference station WBS). HypoDD locations by \citet{liu2020rapid} are shown as black open circles. b) Rectilinearity distribution measured at station WBS. d) Cumulative distance error measured in metres with respect to the HypoDD locations}.
    \label{fig:ridge_auto}
\end{figure}

Fig. \ref{fig:ridge_auto} shows the final result obtained with HADES-R, after automated cluster rotation. In the horizontal plane, the cluster is well reconstructed. In cross-sections, the depth of the events is deeper with respect to the HypoDD locations. However, the performance is comparable to the result with 8 master events in Fig. \ref{fig:ridge_all}. From the rectilinearity calculated at station WBS, we can observe that the majority of the events follows a linear relationship between distance and differential arrival time, with the exception of 6 outliers. 

This benchmark test shows that when many (well-located) master events are available, it remains preferable to use HADES. However, when only few master events are available, using HADES-R with one master event should be preferred for a better reconstruction of the general shape and orientation of the cluster.

\rhead{Data}
\section{Data}

After the successful benchmark test in the previous section, HADES-R will be applied to real data examples in a single borehole DAS monitoring configuration.

\subsection{The Utah FORGE 2019 pilot stimulation}
 
\begin{figure}[H]
    \centering
    \includegraphics[width=\textwidth]{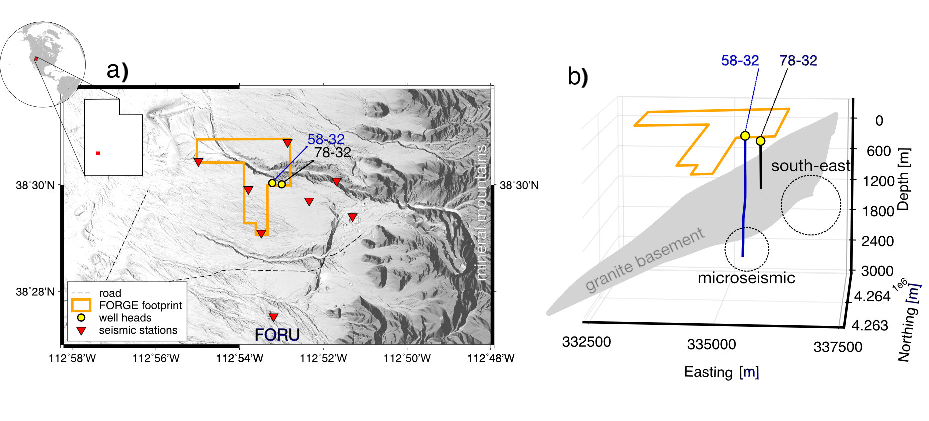}
    \caption{a) Map of the FORGE EGS site, Utah, USA. The surface seismic network in place during the 2019 pilot stimulation is shown with red triangles. The stimulation (58-32) and monitoring (78-32) wells are indicated by yellow circles. b) 3-D rendering of the FORGE footprint highlighting the dipping granite basement interface (grey surface), the pilot injection well 58-32 (blue line) and the monitoring well 78-32 instrumented with DAS (black line). The label "microseismic" indicates the zone around the injection location during the 2019 stimulation, whereas the label "south-east" indicates a region from which (micro-)seismic events unrelated to the EGS activities were recorded during the 2019 stimulation.}
    \label{fig:data_FORGE}
\end{figure}

The Utah FORGE (Frontier Observatory for Research in Geothermal Energy) geothermal test site is an underground field laboratory sponsored by the U.S. Department of Energy, whose purpose is to develop, test, and accelerate breakthroughs in Enhanced Geothermal Systems (EGS) in a controlled environment. FORGE is located in a low-risk area, a few kilometres north of the small town of Milford in Beaver County, Utah, on the western flank of the Mineral Mountains (Fig. \ref{fig:data_FORGE}; \citep{moore2019utah, wannamaker2020geophysical, pankow2019micro}). FORGE is located on a sediment-filled basin with thickness of the sediments between 300 - 1000 m, until they are cross-cut by an east-west sloped granite basement interface. The granite cross-cuts the DAS-instrumented monitoring well at ~800 m depth from the surface. The target EGS reservoir is to be created in the basement granitoid rocks where temperatures have been estimated to exceed 190° C at less than 3 km depth (Fig. \ref{fig:data_FORGE}). The site is constantly evolving as new wells are being drilled and new instrumentation is being incorporated into its monitoring infrastructure. 

To test HADES-R, we focus both on synthetic data and on DAS data collected during phase 2C of the FORGE pilot stimulation, which took place from 21 April until 3 May 2019 \citep{moore2019utah} in well 58-32. The first hydraulic stimulation was performed in the open hole section at the base of the well (phase I) followed by two stimulations targeting the cased section of the well with the aim to determine the viability of stimulating fractures behind the casing (phase II and III). In 2019, the surface monitoring consisted of five seismic broadband stations and three strong-motion accelerometers (red triangles in Fig. \ref{fig:data_FORGE}), and one station placed in a shallow borehole of ~300 m. A temporary high-density nodal array was also installed, but it was not used in this study. Additional details about the surface monitoring network can be found in previous studies \citep[e.g., ][]{pankow2017local,pankow2019micro} and on the Earthscope website (https://ds.iris.edu/mda/UU/). The surface monitoring was complemented by a monitoring borehole (well 78-32) drilled to a depth of ~1000 m with the last 200 m in the granite, and located ~400 m southeast of the stimulation well 58-32 (Fig. \ref{fig:data_FORGE}). Well 78-32 was equipped with an 12-level geophone string deployed by Schlumberger, and a fibre-optic cable in a metal tube, cemented behind the casing by Silixa. The geophones were positioned at depths of 645-980 m, with 30.5 m spacing between each sensor, and they were recording only during the three stimulation phases. The installed fibre cable included two Multi Mode (MM) fibres, two Single Mode (SM) fibres and a specialized Silixa Constellation fibre. The cables were interrogated using the Silixa Carina system. DAS data was acquired with a 1.02 m channel spacing using a 10 m gauge length, and a sampling rate of 2 kHz for a 1000 m long cable. 
For the synthetic tests, we used the same network configuration that was in place during the 2019 stimulation with a randomly generated cluster, and with arrival times generated with a 1D velocity model derived from \citet{zhang2021high}. During the stimulation, the DAS recorded both the near-well microseismicity caused by stimulated fractures at the toe of stimulation well 58-32 (microseismic zone in \ref{fig:data_FORGE}), and distant low-magnitude seismicity located roughly 2 km to the south-east of well 78-32, hereafter referred to as the "south-east cluster" (south-east zone in Fig. \ref{fig:data_FORGE}). Of the near-well microseismicity, over 400 events were catalogued by Schlumberger using the geophone chains, and just over 110 were detected with DAS by \cite{lellouch2020comparison}. This subset of 110 events was employed in the real-data HADES-R application. The exact location procedure as well as the location uncertainty determination process for the events located with the geophones were not disclosed by Schlumberger, although the spatial uncertainties are published with the catalogued locations. The temporary nodal surface nodal array that was active during the stimulation recorded only 7 of the microseismic events, highlighting the importance of downhole sensors. Of the distant south-east cluster, 82 seismic events were recorded by the DAS and catalogued and located by \cite{lellouch2021low}, with estimated distances between 2-5 km south-east of the stimulation well 58-32. Out of these 82 earthquakes, 39 events had clearly visible P- and S-phase arrivals and thus were used for the second real-data application of the HADES-R approach. Picks were made only for the lower section of the DAS where the signal was consistently good, from ~500 m  to 850 m depth along the DAS cable.

The inter-event distance calculation scales exponentially with the amount of events in the cluster (by $n(n-1)/2$), the computation of one iteration of HADES-R amount from seconds to minutes on a laptop, i.e., for the 320 events in the Ridgecrest example, the CPU time was 336 s for one HADES-R iteration; while for the 39 events in the later tests on the south-east cluster, the CPU time was 7 s (on a 2.3 GHz Quad-Core \textit{Intel} Core i7).

\subsection{DAS data pre-processing}

\begin{figure}[H]
    \centering
    \includegraphics[width=\textwidth]{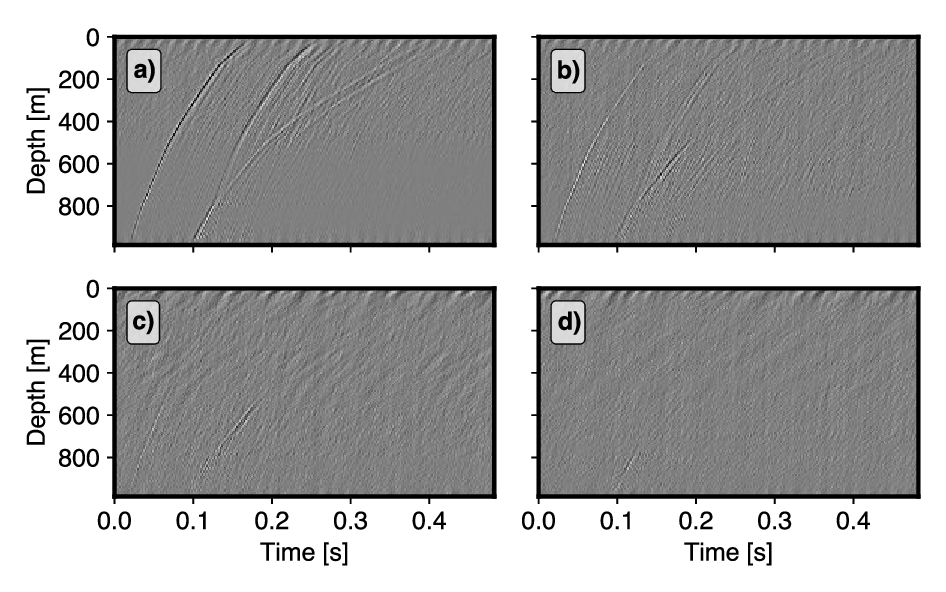}
    \caption{Examples of four types of microseismic event quality from the 2019 DAS microseismic catalogue by \citet{lellouch2020comparison} at FORGE. a) Best quality event: the moveout of the P-wave ($0.02$ - $0.15$ s) and S-wave ($0.1$ - $0.4$ s) are clearly visible across all DAS traces. b) Medium quality event: the arrivals of the P- and S-phases are more difficult to distinguish, yet still visible across almost all traces. c) Poor quality event: The P- and S-phases are only visible on a small section along the DAS cable. d) Very poor quality event: no section of the traces shows a coherent P- and S-phase signal.}
    \label{fig:methods_evtypes}
\end{figure}

Fig. \ref{fig:methods_evtypes} shows four examples of microseismic events of varying magnitudes from the near-well induced microseismicity recorded during the 2019 FORGE pilot stimulation (Fig. \ref{fig:data_FORGE}). We pre-process the raw DAS data by applying the following simple signal analysis: First, we bandpass filter in the frequency range of interest, that is between 10-300 Hz for the microseismic events of cluster A, and between 5-250 Hz for the distant earthquakes of the south-east cluster. Second, we mute noisy or defective channels as well as the shallowest 200 channels due to surface noise, then we apply trace normalisation to enhance first arrival visibility, and a frequency-wavenumber (FK) filter to remove zero-wavenumber and acausal signals, such as down-going reflections. Finally, we apply a 2D median filter. Note how event type d, even after denoising, is too small to adequately pick any phase arrival time (Fig. \ref{fig:methods_evtypes}).

\begin{figure}[H]
    \centering
    \includegraphics[width=0.8\textwidth]{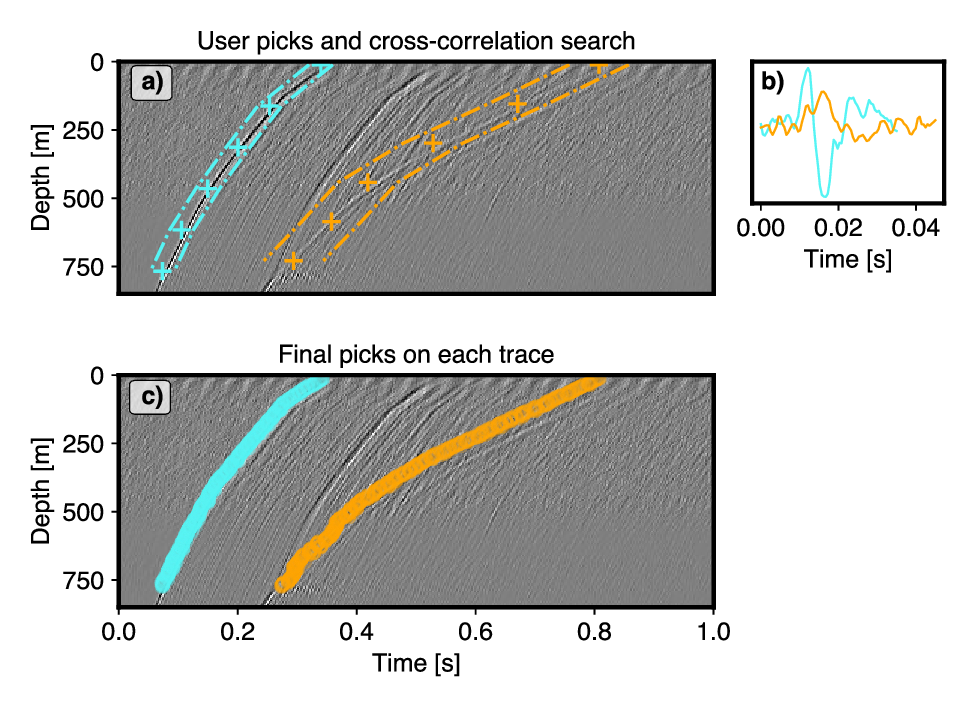}
    \caption{Semi-automated workflow of the proposed picking algorithm for DAS data. a) Hand picking of P-wave (light blue crosses) and S-wave (orange crosses) arrival times and determination of a cross-correlation range (search zone; dotted lines). b) Master trace derived from stacked hand-picked traces over a pre-defined time window, c) Final picks for all traces after a cross-correlation search of the P- and S-templates in the search zone.}
    \label{fig:methods_picking}
\end{figure}

Input data for HADES-R consists of P- and S-wave arrival times. Although picking for conventional seismic (velocity) data is a mature field with various code implementations readily available, these techniques often prove inefficient when used with the large volume of data recorded by DAS, and can fail for very weak (micro-)seismic events due to their poor signal-to-noise ratio on individual channels. For example, STA/LTA pickers cannot be uniquely tuned to accommodate the largely different magnitudes, amplitudes and signal-to-noise ratios found in the FORGE data. Likewise, more sophisticated pickers such as e.g., Phasenet \citep{zhu2019phasenet} and other emerging machine-learning detectors and locators such as MALMI \citep{shi2022malmi} perform well for conventional seismic data, but often require three-component data and extensive labeled training datasets. \citet{lellouch2020comparison} propose to use the estimated angle of arrival of the seismic energy at the bottom of the DAS cable at FORGE, using the velocity profile along the vertical cable to estimate arrival times of the phases at every channel. This efficiently uses the coherency observed in the DAS data. To pick the arrival times in DAS data, we here propose a semi-automated procedure. P and S-wave arrivals are visually picked on traces with high signal-to noise ratio (Fig. \ref{fig:methods_picking}a). In a second step, a user-defined search range is chosen (Fig. \ref{fig:methods_picking}b) that should cover the area around the first arrivals, and this is mostly a wider search area for the S-phase arrival. Finally, the P- and S- master traces (templates) derived from stacking the manual picks are cross-correlated within the previously defined search range to automatically find the picks of all the traces between the manual picks (Fig. \ref{fig:methods_picking}c). 

\rhead{Results}
\section{Results}
In this section, the HADES-R methodology will be applied to the Utah FORGE 2019 pilot test scenario \citep{moore2019utah}. First, a synthetic test in with that test geometry will be used to show how the method performs in the single-borehole DAS setting. Subsequently, results using the real data will be shown for the microseismic monitoring data during injection (microseismic region in Fig. \ref{fig:data_FORGE}), and seismic events taking place 2-5 km south-east of the FORGE network during the pilot test period (south-east region in Fig. \ref{fig:data_FORGE}). Finally, a comparison will be made with results by \citet{lellouch2021low} from the south-east cluster.

\subsection{Synthetic test}
\begin{figure}[H]
    \centering
    \includegraphics[width=0.8\textwidth]{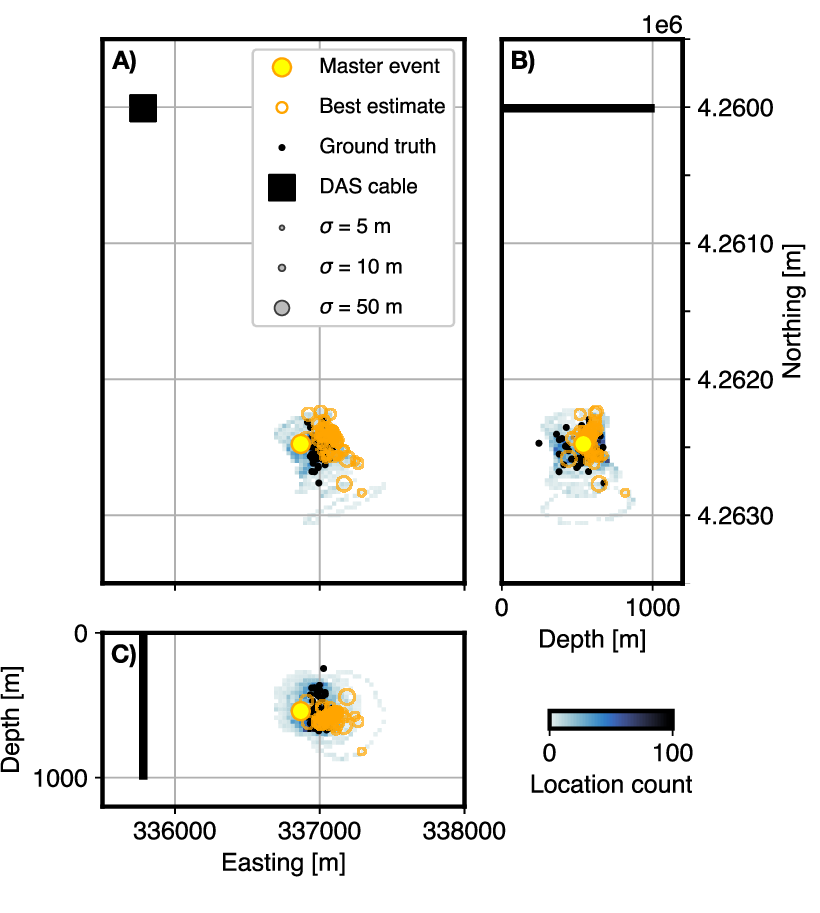}
    \caption{Synthetic test results after 100 channel pair iterations in horizontal (a), longitude-depth (b) and latitude-depth view (c). The optimal location and rotation angle of each channel pair produces a cluster that goes into the blue 2D histogram. It indicates the frequency of event occurrences in each bin and shows the rotational uncertainty. Orange circles indicate the 'best estimate': one iteration solution for which the highest rectilinearity is found. The size of the circles shows the event standard deviations calculated over the 100 iterations.}
    \label{fig:synth_EQ}
\end{figure}

To test the single-well configuration performance with DAS and HADES-R, we first conducted a synthetic test reproducing the FORGE network configuration. We generated a randomly distributed cluster with a slight northeast-southwest striking elongation to represent a more realistic seismic cluster. In all the following tests, depth 0 indicates the wellhead of the DAS-instrumented borehole 78-32. The forward synthetic arrival time dataset was computed using an Eikonal solver by \citet{zunino2023hmclab}, the 1D S-wave velocity model at the location closest to well 78-32 by \citet{zhang2021high}, and a constant $v_p$/$v_s$ ratio of $1.9$. For locating the events, we used the averaged velocities between 200-1200 m depth: $v_p = 3800$ m/s and $v_s = 2000$ m/s. We start from 200 m depth because the true DAS data only had surface noise in this section. Fig. \ref{fig:synth_EQ} shows the obtained locations after 100 iterations of HADES-R and after automatic rotation optimisation has been applied. Random channel pairs with a minimum aperture of 500 m were used. In this plot and all the following plots, a two-dimensional (blue) histogram contains the combination of all final (optimal) cluster location and orientations for the channel combinations. This acts as an uncertainty measure: The cloud shows how well the orientation of the cluster is constrained over different channel combinations and thus angular configurations. The results from Fig. \ref{fig:synth_EQ} show some rotational uncertainty, but a relatively good reconstruction of the shape of the cluster. In the horizontal plane (Fig. \ref{fig:synth_EQ}a), the shape of the cluster is well-reconstructed, and close to the ground truth shape. However, in depth (Fig. \ref{fig:synth_EQ}b and c), HADES-R does not reconstruct the full depth range of the cluster, but the location histogram does give a good representation of the cluster depth range. The 'best estimate' solution is a single HADES-R cluster realisation with the highest rectilinearity score. It is worth noting that the best estimate is mainly useful to show what the computed structure of the cluster looks like, and it always needs to be analysed together with the location cloud. For the best estimate of the synthetic cluster (orange circles in Fig. \ref{fig:synth_EQ}), the size of each circle represents the spatial standard deviation of each event over 100 iterations. The events further away from the master event show larger standard deviations than the events closer to the master event. This can be attributed to the fact that the master event is the origin of the rotational search. 

\subsection{Real data application (1): near-well stimulated microseismicity}
\begin{figure}[H]
    \centering
    \includegraphics[width=\textwidth]{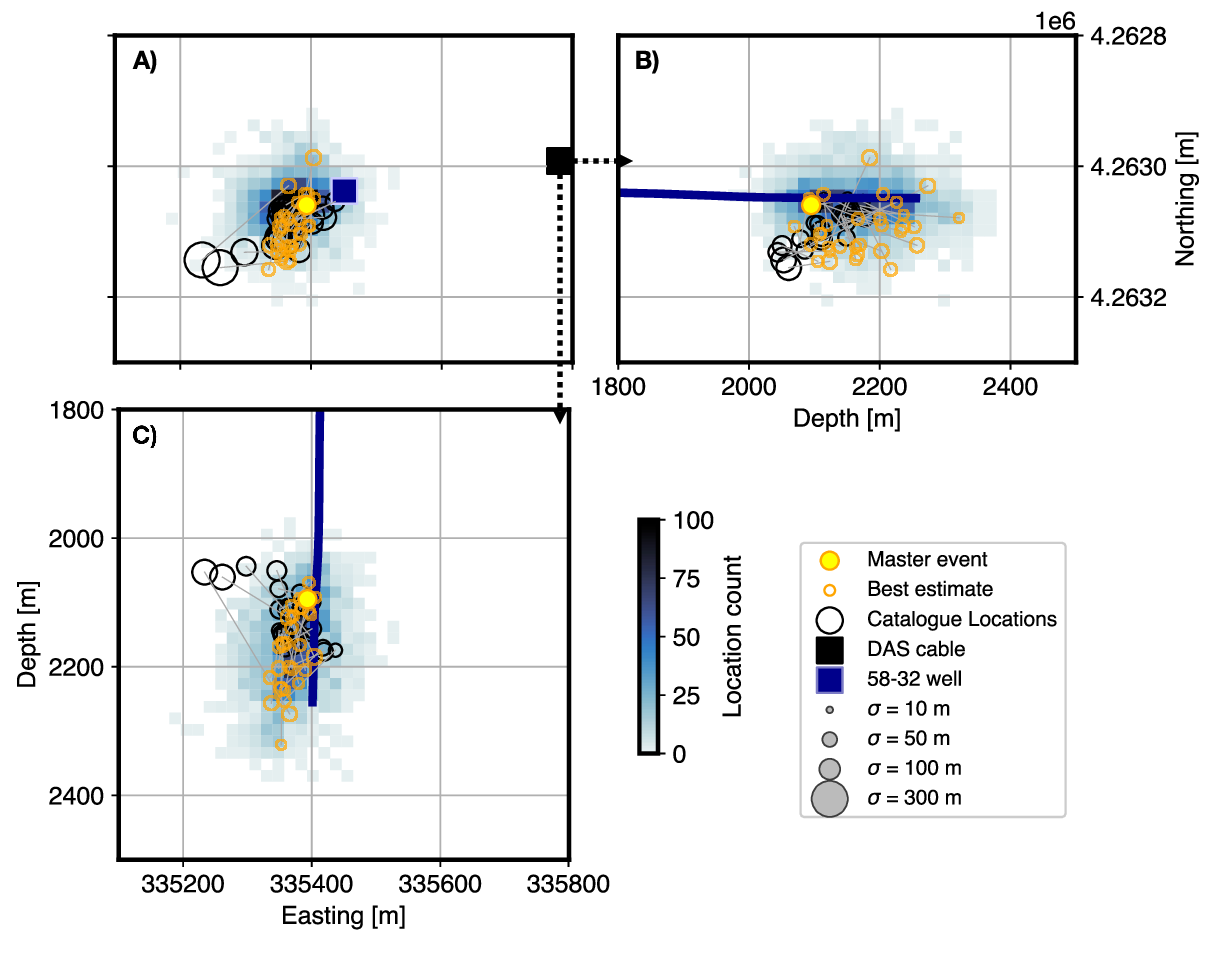}
    \caption{HADES-R results for the near-well microseismicity recorded with DAS during the 2019 FORGE stimulation in the horizontal (a), longitude-depth (b) and latitude-depth view (c). The optimal outcome of 100 channel pair iterations is shown as a two-dimensional histogram (blue colours). Black circles indicate the locations catalogued by Schlumberger using the geophone chain only. The size of the black circles corresponds to the diameter of the major error axis of the error ellipse published with the catalogue. The size of the orange circles corresponds to the standard deviation of the event location over the 100 HADES-R iterations. Lines are drawn between corresponding events from the best estimate HADES-R  result and the catalogue location. Zero depth in this Fig. corresponds to the well head ($\sim$1650 m.a.s.l.). The blue thick line in the figure is the trajectory of the injection well (58-32), while the monitoring well (78-32, black square in panel a) ends at a depth above the range shown in the panels b and c. }
    \label{fig:results_micro}
\end{figure}

Using HADES-R, we located the near-well stimulated microseismicity which was recorded during the FORGE 2019 pilot stimulation (microseismic zone in Fig. \ref{fig:data_FORGE}). Of the 110 catalogued events \citep{lellouch2020comparison}, 32 events were suitable for picking. The location of the chosen master event was taken from the catalogue published by Schlumberger for which the location was derived using the downhole geophone string. As we use a homogeneous velocity model, a challenge could be posed by the basement granite interface that the ray paths between the events and the receivers would intersect at around 800 m depth. For a relatively small cluster (~400 m in width), the ray paths from the events to the DAS channels will not differ significantly, thus we are confident that in this case the granite layer will not substantially affect the shape of the cluster. We caution that the inter-event distance approximation can be over- or under-estimated if the assumptions for the homogeneous velocities are far from the true velocities. Here, we selected velocities of $V_p = 5800$ m/s and $V_s = 3000$ m/s, representing the average velocities in the granite layer, extracted from the 1D velocity model closest to the monitoring well 78-32 \citep{zhang2021high}. We chose 600 m as the minimum depth, because the data quality for some events was too low to allow for clear picking of the shallower channels. Similarly to the synthetic test, the location process is conducted for 100 iterations among random channel pairs for which a minimal aperture of 300 m is set to maximise the array aperture. This aperture is smaller than the previous synthetic test, because only the lower portion of the DAS channels could be picked for a significant part of the microseismic data. The location histogram in Fig. \ref{fig:results_micro} shows a circular pattern in the x-y plane. This is indicative of a high rotational uncertainty, in which the rotation optimiser finds a maximal rectilinearity in many directions. This is likely due to the single-well configuration and lack of azimuthal coverage. However, in depth, the cluster shows less rotational uncertainty, and defines a clear preferred alignment. In Fig. \ref{fig:results_micro} we also overlay, for the same events located with HADES-R, the corresponding catalogued locations obtained by Schlumberger. Schlumberger, for their locations, used only the 12-level geophone string which was installed in the same well as the DAS cable. Due to the challenging configuration between the monitoring well and events, they report location uncertainties on the order of 400 m \citep{moore2019utah}. This is indicated in Fig. \ref{fig:results_micro} by the size of the circles of the catalogue locations, which represents the diameter of the major axis of their associated error ellipses. We observe that the location density distribution obtained with HADES-R encompasses the same area as the catalogue locations. However, our resulting locations for the deeper events appear to be located closer to the injection well, rather than away from the injection well, as one may expect in hydraulic fracturing. This can be attributed to a bias caused by the poor angular configuration of the monitoring well with respect to the injection well. Supporting this hypothesis are the errors of the catalogue locations, which show that the events further away from the injection well have the largest uncertainties, in the order of 300 m. Similar results were also obtained by \citet{lellouch2020comparison}. In their work, they locate in a distance-depth sense (without azimuth) a subset of 45 microseismic events that were detected with DAS in the time window of 27 April 2019, 5 p.m. and 28 April 2019, 5:10 p.m. (UTC). Their DAS-derived locations are also biased towards the monitoring well in the distance-depth plane.

\subsection{Real data application (2): south-east seismic cluster}
\begin{figure}[H]
    \centering
    \includegraphics[width=0.82\textwidth]{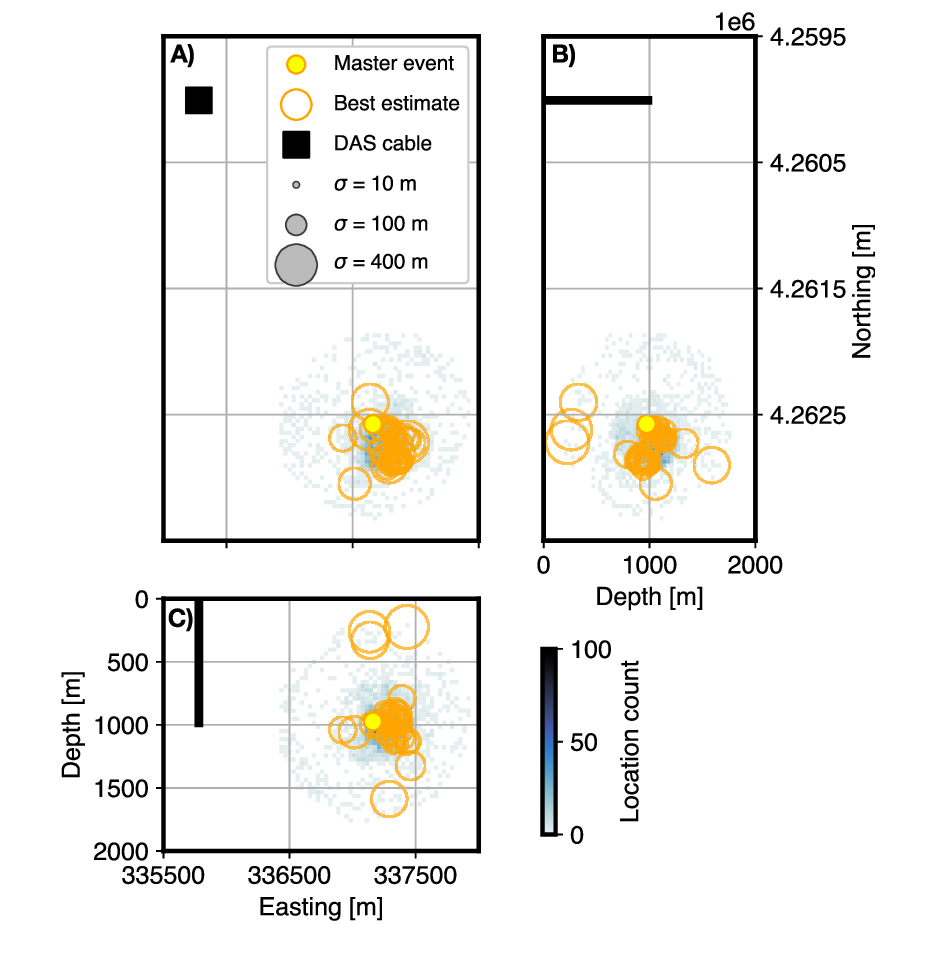}
    \caption{HADES-R results for the distant south-east seismic cluster recorded with DAS during the 2019 FORGE stimulation in the horizontal (a), longitude-depth (b) and latitude-depth view (c). Locations are generated from DAS channel pairs only. The location count histogram indicates the frequency of event occurrences at each specific location in the gridded solution space over 100 iterations in which different station combinations are used. Zero depth in this Fig. corresponds to the well head ($\sim$1650 m.a.s.l.)}
    \label{fig:results_EQ_FULL}
\end{figure}

In phase 2C of the FORGE 2019 pilot stimulation, 82 low-magnitude earthquakes were detected by DAS at $\sim$2-5 km south-east of the monitoring well 78-32 (south-east zone in Fig. \ref{fig:data_FORGE}).Out of the 82 earthquakes catalogued by \cite{lellouch2021low}, 39 had clearly visible P- and S-phase arrivals that could be picked on the lower 626 m section of the DAS cable. One event was also visible in the surface network recordings, thus we use this well-recorded event as master event. Its location was constrained using the regional seismic stations by \citet{lellouch2021low}, with a fixed depth at 1 km during the location inversion, due to the persistently high uncertainties in depth estimation. As in the previous real-data example, our location procedure was repeated over 100 random channel pairs along the DAS cable. We chose the velocity values of $V_p = 5000$ m/s, $V_s = 2600$ m/s as inputs for the velocity model. These values represent the average of the seismic velocities to a depth of 200-2000 m from the well head of the monitoring well 78-32 (closest 1D model by \citet{zhang2021high}). The histogram shows a high rotational uncertainty around the master event (yellow circle in Fig. \ref{fig:results_EQ_FULL}), which is indicative of multiple possible rotations that would maximise the rectilinearity. Nevertheless, there is a noticeable tendency for the events to cluster primarily to the southeast of the master event. This tendency is also reflected in the best estimate locations of different rotation combinations, for which the highest rectilinearity was found. The observed high rotational uncertainty may be caused by  picking errors, incorrect homogeneous velocity estimates, or by inaccuracy of the master event depth, which had to be constrained at 1 km.

\begin{figure}[H]
    \centering
    \includegraphics[width=0.75\textwidth]{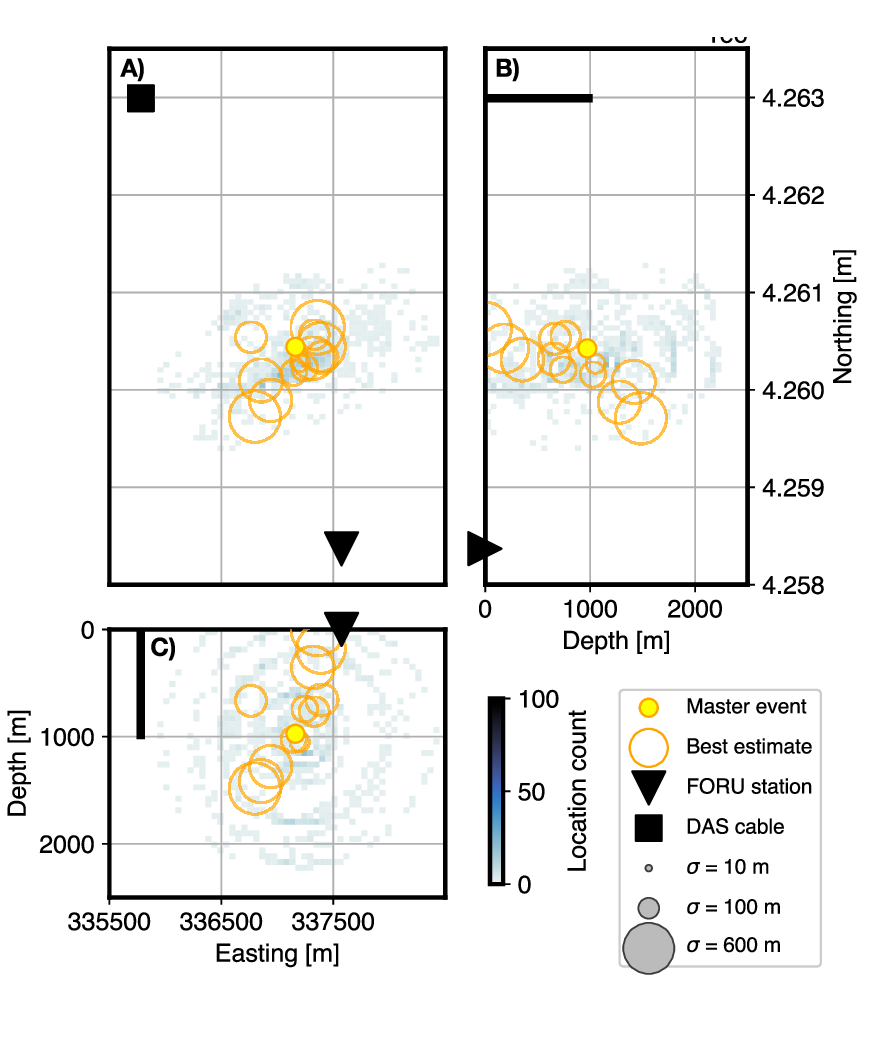}
    \caption{HADES-R results for a subset of events of the distant south-east seismic cluster recorded during the 2019 FORGE stimulation in the horizontal (a), longitude-depth (b) and latitude-depth view (c). Locations are generated by using station FORU and different DAS channels. The optimal outcome of 100 channel pair iterations is shown as a two-dimensional histogram (blue colours). Zero depth in this Fig. corresponds to the well head ($\sim$1650 m.a.s.l.).}
    \label{fig:results_EQ_SHORT}
\end{figure}

To determine if we can reduce the uncertainties in case of additional azimuthal coverage, we locate a subset of the events belonging to the south-east cluster using also the FORU station from the University of Utah surface seismic network (Fig. \ref{fig:data_FORGE}). The FORU station together with the DAS channels forms a large-aperture receiver array, which is beneficial for the angular criteria of HADES-R. FORU also had a better signal-to-noise ratio for these events with respect to the other stations. Of our previously located 39 events, 13 events had clearly visible P- and S-phase arrivals at FORU. Similarly to the previous cases, we perform 100 iterations with station pair combination that include FORU and 100 channels of the DAS cable located in the deeper section (>500 m) of the cable. The results in Fig. \ref{fig:results_EQ_SHORT} highlight how we are able to better constrain the locations in the horizontal directions, as the rotational uncertainty decreases and the circular shape is replaced by a south-west to north-east striking alignment. Moreover, the vertical distribution of events is now slightly better constrained in terms of Northing, and,in depth, events are located within a slightly larger depth range (0-2000 m). 

\begin{figure}[H]
    \centering
    \includegraphics[width=\textwidth]{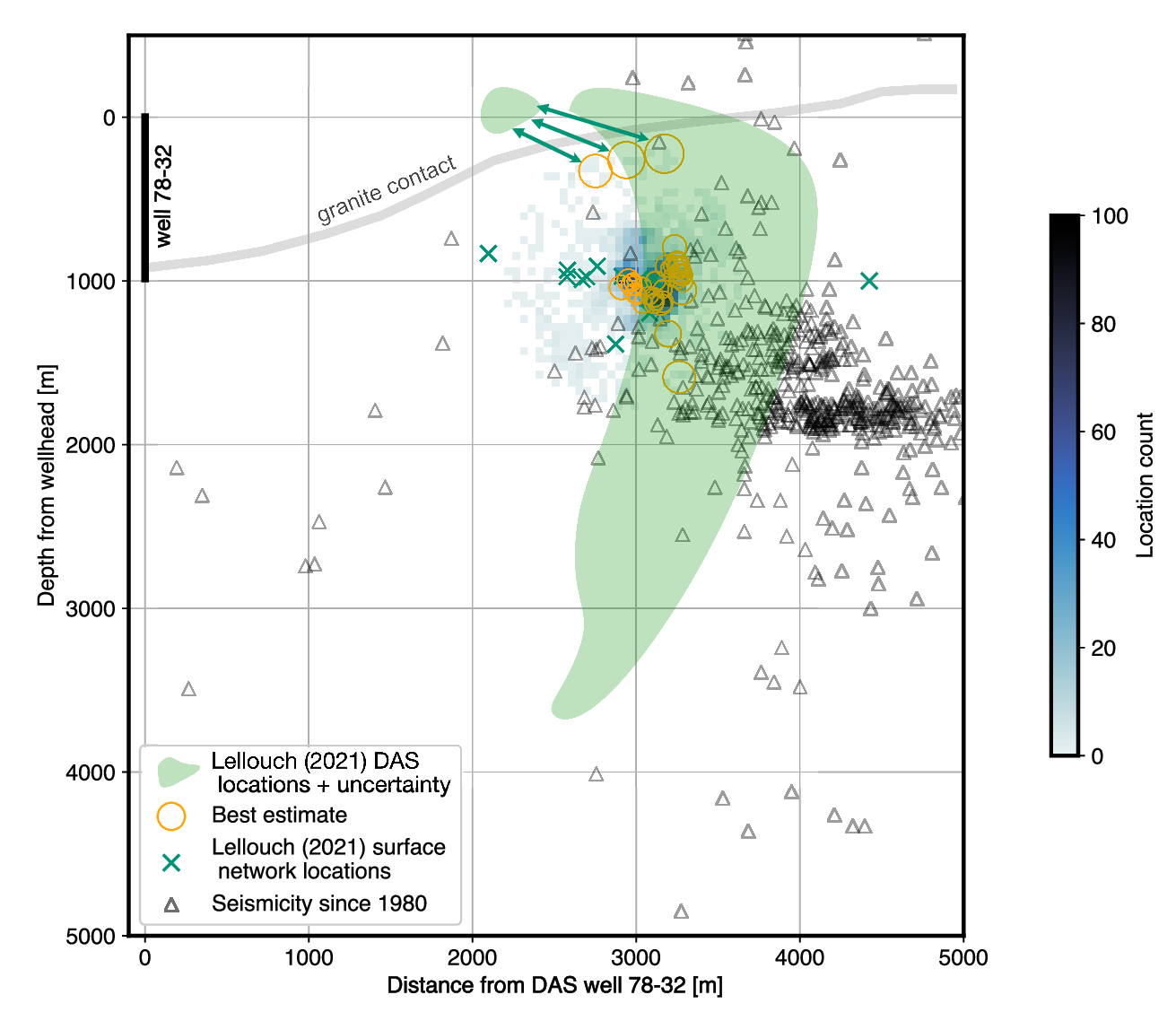}
    \caption{HADES locations for the 2019 south-east seismic cluster compared with historical seismicity for the period between January 1981 and March 2023 catalogued by the University of Utah Seismograph Stations (black triangles). The green patch shows the outline of the DAS depth-distance results obtained by \cite{lellouch2021low}, whereas the green crosses indicate the events located by \citet{lellouch2021low} using the surface network array. The thick grey line shows the approximate granite interface location.}
    \label{fig:disc_overlay}
\end{figure}

Our results tend to agree with the DAS-derived locations of \cite{lellouch2021low}. In Fig. \ref{fig:disc_overlay} we compare our south-east cluster results obtained with only DAS, to the outline of the area in which \cite{lellouch2021low}'s locations lie, including their uncertainty range (green patch). To compare these results, we transformed our coordinates of events and the histogram to distance from well 78-32. \citet{lellouch2021low} also calculated, using the surface network array, the locations for a subset of events that were visible in the surface data (green crosses in Fig.\ref{fig:disc_overlay}). These locations have a horizontal location uncertainty of ~1.5 km and a vertical uncertainty range of ~3 km. One of these events was used as the master event in our HADES-R location procedure. Also plotted is the historical seismicity recorded since 1981 and the HADES-R location results of the south-east cluster (best estimates and location density distribution), generated from DAS channel pairs only. From the comparison we observe that the depth distribution of our locations show a smaller uncertainty density cloud whereas locations by \citet{lellouch2021low} define a much wider area and a greater uncertainty range. Our smaller depth range could be associated with the homogeneous velocity model that is used by HADES-R, that does not account for the granite layer interface and thus could underestimate the velocity of deep event arrivals, while overestimating the velocity of shallow event arrivals. Another contributing factor could be the unknown depth of the chosen master event, which was constrained at 1 km depth, but had 3 km uncertainty. 

An illustration of these potential effects is provided by three events (Fig. \ref{fig:disc_overlay}), that in our HADES-R best estimate results (indicated by arrows) are located at shallower depths than the rest of the cluster, which locates just below the lower end of the DAS cable (985 m). These events have been identified by \cite{lellouch2021low}, due to their lack of a strong S-P conversion, as events refracting off the granite contact, and occurring above the granite interface. Thus, our location estimates are likely too deep, potentially due to the homogeneous velocity assumptions required by HADES-R, or the depth of our master event that was constrained at 1 km. Nonetheless, a clear separation and the a similar orientation of these events is found, and the location results from HADES-R and of \cite{lellouch2021low} both show a strong correlation with the historical seismicity underlying the Mineral Mountains. 

\rhead{Discussion \& Conclusion}
\section{Discussion and Conclusions}
This study introduces several advancements of the relative earthquake location procedure HADES \citep{grigoli2021relative} for clustered seismicity, resulting in the new HADES-R. HADES-R benefits specific seismic location problems: clustered seismicity that is occurring out-of-network and is either recorded by very ill-posed or sparse networks, that do not allow for good absolute location estimates. 

Prerequisites are that the P- and S-phases are recorded at a minimum of two receivers and that at least one event in the cluster has a well-known absolute location (the master event). Additionally an estimate of the cluster width, and a homogeneous velocity model estimate are required. Examples of use cases are offshore sequences that have been recorded by on-land receivers, sparsely instrumented areas such as volcanic environments or glaciers, a seismic cluster occurring outside of a borehole network, a single-borehole seismic network, or long-term monitoring with a rudimentary seismic network after hydraulic stimulations. In the case that one or a few events in such seismic sequences are large enough to absolutely locate with a larger array of sensors, or if there was a check- or perforation-shot in hydraulic stimulation applications, or if earlier denser instrumentation already enabled locating events from the same cluster, we can apply HADES or HADES-R. The choice between the two depends on the amount of master events, and the uncertainty of their locations. HADES-R requires a single master event, whereas the older HADES can be used upwards from 4 master events and results improve with increasing master event numbers. However, if there are four or more (potentially poorly) absolutely located events, it may still be beneficial to use HADES-R to avoid a erroneous cluster shape due to large master event uncertainty, as demonstrated in our benchmark test. 

For sparse network scenarios, we conducted a benchmark test using the Ridgecrest 2019 \citep{liu2020rapid} dataset. This test showed that it is possible to determine the shape and orientation of a seismic cluster with one master event and two near-orthogonal oriented stations. Multiple single master event iterations of HADES-R even outperformed HADES with only a few (5-8) available master events, when compared to HypoDD locations obtained with a much denser (30+ stations) seismic network. 

We do not only show that clustered seismicity can be located efficiently using very little prior location information, we also show the application for DAS-only network configurations in a single-borehole monitoring setting. Locating events using DAS without supplementary instruments has gained interest in microseismic monitoring. Particularly in environments where geophones may mechanically fail due to factors such as high temperatures, or where surface noise impedes the detection of deep small-magnitude events \citep{binder2021joint}. The numerous channels in DAS are effectively utilised in HADES-R. Along with the azimuthal information derived from a pre-located master event, they help compute a cluster orientation uncertainty cloud. This cloud contains the locations from iterations of HADES-R over different channel pairs and illustrates the resulting cluster shape and orientations. Large uncertainties in shape and orientation are reflected in wide, round location clouds, whereas a more constrained cloud shows consistent locations over all iterations. We note that, in a single-borehole DAS configuration,  we are still quite limited in resolving the true azimuthal ambiguity in the x-y plane. By using the the differential traveltime for estimating the inter-event distances, the third dimension (cluster width) has to be input and estimated by other means (e.g., geological constraints, waveform clustering approaches). However, we can reduce the depth uncertainty. By assuming that all located events belong to the same cluster, and that the distance between the receivers and the cluster centroid is much larger than the overall size of the cluster, we can thus consider the locations obtained with HADES-R a reasonable approximation of the true locations. This assumption can be especially valid in microseismic monitoring applications, where a constrained cluster around the injection zone is expected. 

Considering the FORGE 2019 single-borehole setting as an example, we have first shown with a synthetic test that, even when we use a 1D forward model , we can recover the shape of the cluster. The real data examples have shown that, although the microseismic cluster as well as the south-east cluster can be reasonably located, the uncertainties remain large and no clear cluster orientation is found. This can be attributed to the fact that the optimal conditions for HADES-R are not satisfied in either scenario: in the microseismic cluster setting, the angle between the channel pairs and the cluster is never close to 90\degree, and in the south-east cluster setting, the spread of the cluster is far wider than the distance of the cluster to the stations. Nevertheless, the results are still informative about the shape of the cluster, albeit with ambiguous orientation. Additionally, the results of HADES-R achieved smaller standard deviations (10-50 m) compared to the catalogued locations obtained with the geophone chains. 

In the case of the south east cluster, DAS was the only instrument actively recording, as downhole geophones were not available and most of the surface stations were too noisy to detect these events. Using only the DAS recording, we were able to locate a subset 32 of the 82 earthquakes detected by \cite{lellouch2021low} and obtained standard deviations ranging from 10-400 m. The resulting locations are comparable with historical seismicity and locations obtained by \citet{lellouch2021low} in the depth-distance plane (Fig. \ref{fig:disc_overlay}). Additionally, we combined multiple DAS channels with one surface station to better constrain the azimuthal orientation of the cluster. This approach using a sub-set of visible events on the surface station improved our understanding of the cluster shape. Overall, the single-well monitoring would  benefit from a deeper borehole that extends to the depth at which the events are taking place. This would increase the network aperture and facilitate the retrieval of accurate locations for events occurring further away from the stimulation well. 

We have shown here that HADES-R can provide valuable information about seismic clusters in challenging network conditions. The method's iterative approach, adaptability to different instrument types (e.g., seismometers, geophones, DAS), and ability to constrain cluster shape and uncertainties, offer a promising avenue for monitoring seismic activities in sparse, and ill-posed network configurations. Additionally, the modular framework upon which HADES is built, allows for integration with other potential approaches. This includes adding the capability to iterate over different P- and S-wave speeds, master and reference event combinations, as well as cluster width estimations. Alternative methods that calculate inter-event distances without relying on differential arrival times, can be, for example, easily incorporate into the inter-event distance matrix of HADES-R.
Thus, HADES-R offers a versatile tool that can act as a prior for more advanced location routines. It's ability to reconstruct the general shape and orientation of a seismic cluster and retrieve the distribution of location uncertainty, is a key step forward to advance our capacity to respond to and understand seismicity in environments where traditional tools might fail or resources are limited.

\newpage
\rhead{Other}
\section{Acknowledgements}
We would like to thank FORGE for making all data from the laboratory public in the Geothermal Data Repository, \url{https://gdr.openei.org/forge}. We would also like to thank Olivier den Ouden and Andrea Zunino for discussions that helped improve this paper. This work was supported by the De-Risking Enhanced Geothermal Energy project (Innovation for DEEPs). DEEP is subsidized through the Cofund GEOTHERMICA, which is supported by the European Union’s HORIZON 2020 programme for research, technological development, and demonstration under Grant Agreement Number 731117.

\section{Data Availability}
If this paper is accepted, the HADES-R package will be available for downloading through GitHub. All regional seismic data were downloaded through the IRIS Wilber 3 system (\url{https://ds.iris.edu/wilber3/}) or IRIS Web Services (\url{https://service.iris.edu/}), including the following seismic networks: (1) the UU (University of Utah Regional Seismic Network; 1962). Utah FORGE data is made available via the Geothermal Data Repository: 2019 stimulation DAS data and seismic catalogues can be found at \url{https://dx.doi.org/10.15121/1603679}. We made use of the python package Obspy \citep{beyreuther2010obspy} for our data processing routines. The figures in this paper have been generated using the python library MATPLOTLIB \citep{hunter2007matplotlib} and PyGMT \cite{uieda2021pygmt}. 

\section{Conflict of interest}
This paper has no conflict of interest.

\bibliography{ref}

\newpage
\begin{appendix}

\rhead{Appendix}
\section{Appendix}
\label{sec:supplementary}
\renewcommand{\theequation}{\Alph{section}.\arabic{equation}}
\setcounter{equation}{0} 

\subsection{Inter-event distance calculation: single station case}
\label{sec:sup_iedist}

Given seismic events $a$ and $b$ with coordinates $\mathbf{x}_{a}$ and $\mathbf{x}_{b}$ in $\mathbb{R}^3$, and a \textit{single} seismic station $s$ with coordinates $\mathbf{x}^{s}$, the inter-event distances are:

The distance between events $a$ and $b$ is:
\begin{equation}
\left|\left| \mathbf{r}_{ab}\right|\right|=\sqrt{(x_{a}-x_{b})^2+(y_{a}-y_{b})^2+(z_{a}-z_{b})^2}
\label{eqa1}
\end{equation}
The distance between events $a$ and $s$ is:
\begin{equation}
\left|\left| \mathbf{r}_{a}^{s}\right|\right|=\sqrt{(x_{a}-x^{s})^2+(y_{a}-y^{s})^2+(z_{a}-z^{s})^2}
\label{eqa2}
\end{equation}
The distance between events $b$ and $s$ is:
\begin{equation}
\left|\left| \mathbf{r}_{b}^{s}\right|\right|=\sqrt{(x_{b}-x^{s})^2+(y_{b}-y^{s})^2+(z_{b}-z^{s})^2}
\label{eqa3}
\end{equation}
We aim to estimate $\left|\left| \mathbf{r}_{ab}\right|\right|$ using data from the seismic station. By expressing $\left|\left| \mathbf{r}_{ab}\right|\right|^2$ in terms of $\left|\left| \mathbf{r}_{a}^{s}\right|\right|$ and $\left|\left| \mathbf{r}_{b}^{s}\right|\right|$, we find:

\begin{equation}
\left|\left| \mathbf{r}_{ab}\right|\right|^2=\left|\left| \mathbf{r}_{a}^{s}\right|\right|^{2} + \left|\left| \mathbf{r}_{b}^{s}\right|\right|^{2} - 2\left|\left| \mathbf{r}{a}^{s}\right|\right| \left|\left| \mathbf{r}_{b}^{s}\right|\right|\cos{\phi}
\label{eqa4}
\end{equation}

where $\phi$ is the angle between $\mathbf{r}_{a}^{s}$ and $\mathbf{r}_{b}^{s}$, evaluated in the plane generated by these two vectors (assuming that $\mathbf{r}_{a}^{s}$ and $\mathbf{r}_{b}^{s}$ are linearly independent vectors). If the inter-event distance between $a$ and $b$ is much smaller than the distance between these events and the station $s$ ($\cos{\phi}\approx 1$), we have an approximate estimate:

\begin{equation}
\left|\left| \mathbf{r}_{ab}^{s}\right|\right|=\mid \left|\left| \mathbf{r}_{a}^{s}\right|\right| - \left|\left| \mathbf{r}_{b}^{s}\right|\right| \mid
\label{eqa5}
\end{equation}

Equation (\ref{eqa5}) provides a reasonably accurate estimate of the inter-event distance $\left|\left| \mathbf{r}_{ab}^{s}\right|\right|$ when the station and both events are located along the same line. It is less accurate when the station and both events lie on perpendicular lines. A more accurate inter-event distance can be computed with the use of two stations $S1$ and $S2$, resulting in the Pythagorean inter-event distance calculation:

\begin{equation}
    \left|\left| \mathbf{r}_{ab}\right|\right| \approx  \sqrt{ \left|\left| \mathbf{r}_{ab}^{s1}\right|\right|^{2} + \left|\left| \mathbf{r}_{ab}^{s2}\right|\right|^{2}}
\label{eqa6}
\end{equation}

\subsection{Locating events within the cluster}
\label{sec:sup_clusloc}

\begin{figure}[H]
    \centering
    \includegraphics[width=0.9\textwidth]{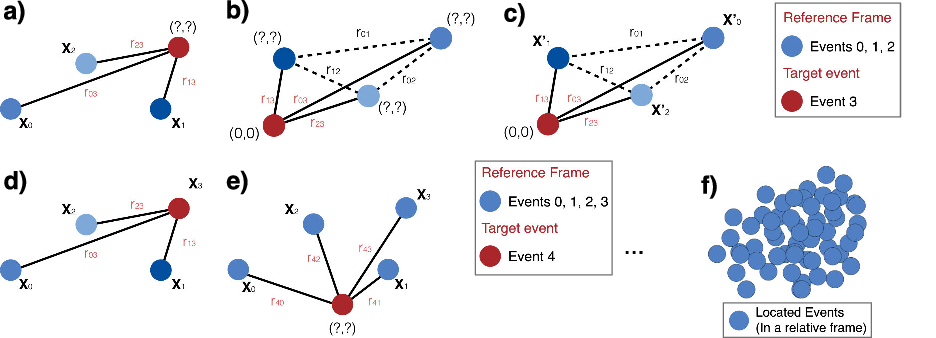}
    \caption{Location procedure of additional events with respect to the reference frame. a) a new target event (Event 3) is to be located with respect to our previously built reference frame. In b) we place the target event at the origin, and in c) given the estimated inter-event distances $r_{ij}$ and the known distances between the reference events, we compute the positions of the already located events relative to event 3 ($\mathbf{X'}$). Then in d) the system is transformed back into its original position and event 3 has become part of the already-located events. Then the next target event 4 is to belocated (e). This process is repeated until all events are located relative to one another (f).}
    \label{fig:methods_evloc}
\end{figure}

The proposed location method requires at least four non-coplanar seismic events,  of which only the absolute hypocentral coordinates of the \textit{master event} ($e_0$) needs to be known. The location of the three \textit{reference events} can be unknown, but there must be an estimate of the width of the cluster. This ensures the uniqueness of the inverse problem. Let the seismicity cluster consist of $k$ events, represented by hypocentral coordinates $\mathbf{x}_{1}, \mathbf{x}_{2}, ..., \mathbf{x}_{k+1}$ in $\mathbb{R}^3$. The $k$ events have already been located (blue events in Fig. \ref{fig:methods_evloc}), and the $k+1$ is the new target event to be located. The euclidean distance between events $i$ and $j$ is denoted by $\left|\left| \mathbf{r}_{ij}\right|\right|$, and the distances between the events and the origin of the reference system are $\left|\left| \mathbf{x}_{i}\right|\right|$ and $\left|\left| \mathbf{x}_{j}\right|\right|$, respectively.

We consider the seismicity cluster's internal structure invariant to translation and rotation, allowing us to place a new event $k+1$ at the origin of a relative coordinate system (i.e., $\mathbf{x}_{k+1}=(0,0,0)$), see Fig. \ref{fig:methods_evloc}a and b. Consequently, $\left|\left| \mathbf{r}_{ik+1}\right|\right|^2=\left|\left| \mathbf{x}_{i}\right|\right|^{2}$ and $\left|\left| \mathbf{r}_{jk+1}\right|\right|^2=\left|\left| \mathbf{x}_{j}\right|\right|^{2}$. This leads to the equation:
\begin{equation}
\left|\left| \mathbf{r}_{ij}\right|\right|^2=\left|\left| \mathbf{r}_{ik+1}\right|\right|^{2} - 2\mathbf{x}_{i}^T\mathbf{x}_{j}+ \left|\left| \mathbf{r}_{jk+1}\right|\right|^{2} \quad \text{with} \quad i,j=1,2,...,k
\label{eqa7}
\end{equation}

The next step is, given the calculated approximated inter-event distances, to find the locations of the already located events $1, ..., k$ relative to the new event $k+1$. The solution to this is to solve a least-squares problem that finds the locations of all events, while trying to fit the approximated inter-event distances. To this end, we define a $k \times 3$ matrix $\mathbf{X}$ containing the coordinates of the first (already located) $k$ events:
\begin{equation}
\mathbf{X}=\mathbf{x}_{i} \quad \text{with} \quad i=1,2,...,k.
\label{eqa8}
\end{equation}

Then, a $k \times k$ distance matrix $\mathbf{R}$ containing the distances of all event pairs is given by:
\begin{equation}
\mathbf{R}=\left( \left|\left| \mathbf{r}_{ik+1}\right|\right|^{2} -\left|\left| \mathbf{r}_{ij}\right|\right|^2 + \left|\left| \mathbf{r}_{jk+1}\right|\right|^{2} \right)/2 \quad \text{with} \quad i,j=1,2,...,k
\label{eqa9}
\end{equation}

 The singular value decomposition of $\mathbf{R}$ is:

 \begin{equation}
    \mathbf{R}=\mathbf{V}\mathbf{S}\mathbf{V}^T ,
\label{eq10}
\end{equation}

we find orthogonal matrices $\mathbf{V}$ and $\mathbf{V}^T$ that contain the eigenvectors of $\mathbf{R}$ and the diagonal matrix $\mathbf{S}$ contains the singular values of $\mathbf{R}$. These can be used to find the event coordinates in $\mathbf{X}$, via solving:

 \begin{equation}
    \mathbf{X}=\mathbf{G}\Sigma^{1/2} .
\label{eq11}
\end{equation}
Here, $\mathbf{G}=\mathbf{V}(1:k,1:3)$ and $\Sigma=\mathbf{S}(1:3,1:3)$, and the notation $l:m$ describes extracting the elements of the matrix along one axis from the $l$-th to the $m$-th positions. Work by \citet{havel1998distance} offers a more detailed mathematical description. This procedure helps locate the first $k$ events in the new reference frame with the event $k+1$ at the origin. We then know the positions of the already located events with respect to event $k+1$ as $\mathbf{X'}_k$ (Fig. \ref{fig:methods_evloc}c), and can compute the correct location of event $k+1$ (Fig. \ref{fig:methods_evloc}d). The process can be iteratively applied to the remaining $n-4$ seismic events to locate all $n$ events within the cluster (Fig. \ref{fig:methods_evloc}e-f).

\subsection{Quaternion rotations}
\label{sec:sup_quat}

For rotations we employ quaternions: whereas complex numbers describe rotations in two-dimensional space, quaternions describe rotations around arbitrary axes in higher-dimensional space, specifically using a four-dimensional representation. Quaternions are used in, e.g., computer graphic simulations and robotics, as they can uniquely describe a rotation and are computationally more efficient than rotation matrices \citep{shoemake1985animating}. The quaternion $\mathbf{q}$ is computed using the three components of an axis unit vector $\mathbf{v}$ and the rotation angle $\theta$:
\begin{equation}
    \mathbf{q}(\theta) = [sin(\theta/2), \quad v_0 cos(\theta/2),\quad v_1 cos(\theta/2), \quad v_2 cos(\theta/2)].
\end{equation}

The use of quaternions in this work is motivated by the arbitrary orientation of the axes between cluster and receiver, and its intuitive representation of rotations. The rotation is in this case performed around three axes, so three separate quaternion operations are applied. The first axis is the Z-axis from the master event to the surface, the second axis is the orthogonal axis, that is a projection of the vector between the master event and a receiver, and the third axis is the cross-product of these two axes to ensure orthogonality. The three axes need to be orthogonal to ensure that rotations are independent: rotation around one axes does not affect an event rotation with respect to another axis.

\end{appendix}

\end{document}